\documentclass[20pt]{article}
\usepackage[english]{babel}
\usepackage[utf8]{inputenc}
\usepackage{wrapfig}
\usepackage{amsmath, amssymb, cancel}
\usepackage{amsthm}
\usepackage{mathtools}
\usepackage{amscd}
\usepackage{bbm, stmaryrd}
\usepackage{yfonts}
\usepackage{multicol}
\theoremstyle{definition}
\newtheorem{teo}{Theorem}[section]
\newtheorem{defi}{Definition}[section]

\newtheorem{Lemma}{Lemma}
\newtheorem{cor}{Corollary}

\usepackage{tikz}
\usetikzlibrary{patterns,matrix,through,arrows,decorations.pathreplacing,decorations.markings,snakes,shadows,shapes.geometric,positioning,calc,backgrounds,shapes.arrows,fit,backgrounds}

\tikzset{anchorbase/.style={baseline={([yshift=-0.5ex]current bounding box.center)}},
int/.style={thick},
  cross line/.style={preaction={draw=white,line width=6pt,-}},
  wall/.style={thin,double,blue},
  middlearrow/.style={postaction=decorate,decoration={markings,mark=at
    position .55 with {\arrow{stealth};}}},
  middlearrowrev/.style={postaction=decorate,decoration={markings,mark=at
    position .55 with {\arrowreversed{stealth};}}},
  ev/.style={shape=rectangle, draw},
  every path/.style={line width=1pt}
}  
\def \myweightstyle {}
\def \myweightx {0.09}
\def \myweighty {0.2}
\newcommand{\mydrawdown}[1]%
  {\draw [\myweightstyle] #1++(0,-\myweighty) -- ++(\myweightx,\myweighty);%
    \draw [\myweightstyle] #1++(0,-\myweighty) -- ++(-\myweightx,\myweighty);}
\newcommand{\mydrawup}[1]%
  {\draw [\myweightstyle] #1 -- ++(-\myweightx,-\myweighty);%
    \draw [\myweightstyle] #1 -- ++(\myweightx,-\myweighty);}

\author{Federico Zerbini}
\date{ }
\makeindex
\title{\huge{\huge{\textbf{Single-valued multiple zeta values in genus 1 superstring amplitudes}}}}

\begin{document}
\maketitle

\abstractname{.~ We study the functions $D_{\underline{l}}$ introduced by Green, Russo, Vanhove in \cite{Green4} in the context of type II superstring scattering amplitudes of 4 gravitons on a torus. In particular we describe a method to algorithmically compute the coefficients in their expansion at the cusp in terms of conical sums. We perform explicit computations for 3-graviton functions, which naturally suggest to conjecture that only single-valued multiple zeta values appear.}

\tableofcontents

\section{Introduction}

The low-momentum expansion of scattering amplitudes in string theories has been considered and extensively studied since the end of the sixties, the pioneer being Veneziano and Virasoro, who studied open and closed bosonic string amplitudes (\cite{Veneziano}, \cite{Virasoro}), respectively.
However, progress in this area has been fairly slow: indeed, only in the last years a fruitful interaction between physicists and mathematicians is speeding up the research and producing amazing advances.

The coefficients of the tree-level amplitude in superstring theory are in general fairly well understood, both for open and closed strings\footnote{A precise description of open and closed string theory is beyond the scope of the present work; we refer the reader to the literature.}, with an arbitrary number of particles (see \cite{Stieb2}, \cite{Stieb1}, \cite{Terasoma2}). In particular, the coefficients are multiple zeta values for open strings and single-valued multiple zeta values for closed strings.

For the genus 1 case the situation is much more complicated. Recently some progress has been made in the study of open string amplitudes in \cite{Matthes1}, where very interesting results with an arbitrary number of particles are obtained. The upshot is that the amplitude can be written in terms of elliptic multiple zeta values, functions depending on $\tau\in\mathbb{H}$ whose coefficients in their Fourier expansion involve multiple zeta values (see \cite{Enriquez},\cite{Matthes2}). In the closed string case, already the first physically meaningful amplitude, involving 4 particles, is far from being understood, despite the great effort spent in the last 15 years in this direction (see \cite{Green3},\cite{Green4},\cite{green1},\cite{Green5},\cite{Green6}). Very little is known for higher genus (see \cite{Green9}, \cite{Green10} for genus 2 and \cite{Mafra} for genus 3).

The present paper is concerned with closed strings in genus 1. Let us see how one gets the amplitude in this case. We want to study, more precisely, the low-momentum expansion of the genus 1 scattering amplitude of four gravitons in type II superstring theory. This means in particular that our strings are closed. Each graviton is labelled by its momentum $p_i$, where $p_i^2=0$ and $\sum p_i=0$, and by a kinematic quantity $\zeta_i$ that we do not define, since we do not need it for what follows. One can define the so called \emph{Mandelstam variables} $s:=-(p_1+p_2)^2$,  $t:=-(p_1+p_4)^2$ and $u:=-(p_1+p_3)^2$. The conditions on the momenta are then equivalent to the fact that $s+t+u=0$.

The amplitude in genus 1 has the form
\[
\textbf{A}_{\zeta_1,\zeta_2,\zeta_3,\zeta_4}(s,t,u)=I(s,t,u) \textbf{R}_{\zeta_1,\zeta_2,\zeta_3,\zeta_4},
\]
where $\textbf{R}_{\zeta_1,\zeta_2,\zeta_3,\zeta_4}$ is a kinematic factor (see \cite{Green4}), and $I(s,t,u)$ is a power series in the Mandelstam variables. $I$ is defined to be the integral over the moduli space of elliptic curves (with respect to the appropriate hyperbolic measure) of the following function, which is itself defined as an integral:
\begin{equation}\label{F}
\iint \exp(\frac{\alpha'}{4}(s(G(z-w,\tau)+G(v,\tau))+t(G(w-v,\tau)+G(z,\tau))+u(G(z-v,\tau)+G(w,\tau)))) \frac{dzdwdv}{\tau_2^3},
\end{equation}
where:
\begin{itemize}
\item $\tau=\tau_1+i\tau_2\in\mathbb{H}$,
\item we perform the integral over 3 copies of the torus $(\mathbb{C}/\Lambda_\tau)^3$, for $\Lambda_\tau=\mathbb{Z}\tau+\mathbb{Z}$,
\item $z,w,v$ represent 3 of our 4 gravitons moving on the torus (without loss of generality we can assume that the fourth is fixed and we choose it to be in the origin),
\item $\alpha'$ is a physical constant (the inverse of the string tension),
\item $G$ (the propagator function) is the Green function on the torus defined as
\[
G(z,\tau):=-\log\bigg|\frac{\theta_1(z,\tau)}{\eta(\tau)}\bigg|^2+\frac{2\pi z_2^2}{\tau_2},
\]
where $z=z_1+iz_2$.
\end{itemize} 
Recall that the Jacobi theta function $\theta_1(z,\tau)$ and the Dedekind eta-function $\eta(\tau)$ are defined by
\begin{eqnarray*}
\theta_1(z,\tau)&=&\sum_{\nu\in \mathbb{Z}+1/2}(-1)^{\nu-1/2}q^{\nu^2/2}\zeta^\nu=-q^{1/8}\zeta^{-1/2}\prod_{n\geq 1}(1-q^n)(1-q^{n-1}\zeta^{-1})(1-q^{n}\zeta),\\
\eta(\tau)&=&q^{1/24}\prod_{n>0}(1-q^n),
\end{eqnarray*}
with $q=e^{2\pi i\tau}$ and $\zeta=e^{2\pi iz}$.

An easy computation shows that the function $G$ is elliptic and modular invariant with respect to the action of $\mbox{SL}_2(\mathbb{Z})$.

To perform the integration one can expand the exponential as a power series in $\alpha's,\alpha't,\alpha'u$, and then one gets linear combinations of functions $D_{\underline{l}}(\tau)$ defined for $\underline{l}=(l_1,\ldots,l_6)\in\mathbb{Z}_{\geq 0}^6$ as
\[
D_{\underline{l}}(\tau):=\frac{1}{4^3}\iint_{(\mathbb{C}/\Lambda_\tau)^3} G(z-w,\tau)^{l_1}G(v,\tau)^{l_2}G(w-v,\tau)^{l_3}G(z,\tau)^{l_4}G(z-v,\tau)^{l_5}G(w,\tau)^{l_6} \frac{dzdwdv}{\tau_2^3}.
\]
The properties of $G$ imply that this integral is well defined, and that $D_{\underline{l}}$ is a modular function. Our aim is to understand better these functions. We call $l_1+\cdots+l_6$ the \emph{weight} of $D_{\underline{l}}$.

It is worth mentioning that one can easily deduce a series representation of $D_{\underline{l}}$ (see \cite{Green4}):
\begin{equation}\label{sum}
D_{\underline{l}}(\tau)=\Big(\frac{\tau_2}{4\pi}\Big)^{l_1+\cdots +l_6}\sum \prod_{j=1}^6\prod_{i=1}^{l_i}\mid\omega^{(j)}_i\mid^{-2},
\end{equation}
where the sum runs over the lattice points $\omega^{(j)}_i:=m^{(j)}_i\tau+n^{(j)}_i$ such that
\[
\omega^{(1)}_1+\cdots+\omega^{(1)}_{l_1}+\omega^{(4)}_1+\cdots+\omega^{(4)}_{l_4}+
\omega^{(5)}_1+\cdots+\omega^{(5)}_{l_5}=0,
\]
\[
\omega^{(3)}_1+\cdots+\omega^{(3)}_{l_3}+\omega^{(6)}_1+\cdots+\omega^{(6)}_{l_6}=
\omega^{(1)}_1+\cdots+\omega^{(1)}_{l_1},
\]
\[
\omega^{(3)}_1+\cdots+\omega^{(3)}_{l_3}+\omega^{(5)}_1+\cdots+\omega^{(5)}_{l_5}=
\omega^{(2)}_1+\cdots+\omega^{(2)}_{l_2}.
\]
However, in the present work we will rather use the definition of $D_{\underline{l}}$ as an integral. Following the method of Green, Russo and Vanhove in \cite{Green4}, we write
\begin{equation}\label{G=B+P}
G(z,\tau)=2\pi\tau_2\,\overline{B_2}(z_2/\tau_2)+P(z,\tau),
\end{equation}
where
\[
P(z,\tau):=\sum_{\substack{k\neq 0\\ n\in\mathbb{Z}}}\frac{1}{|k|}e^{2\pi ik(n\tau_1+z_1)}e^{-2\pi|k||n\tau_2-z_2|}
\]
and $\overline{B_2}$ is the second generalized Bernoulli polynomial, defined as $\overline{B_2}(x)=B_2(x)=x^2-x+1/6$ for $x\in[0,1]$ and then defined imposing $\overline{B_2}(x+1)=\overline{B_2}(x)$ for every $x\in\mathbb{R}$.

Then we have proved the following:
\begin{teo}\label{main}
For every $\underline{l}=(l_1,l_2,l_3,l_4,l_5,l_6)$ we have
\[
D_{\underline{l}}(\tau)=\sum_{\mu,\nu\geq 0}d_{\underline{l}}^{(\mu,\nu)}(\pi\tau_2)\,q^\mu\,\overline{q}^\nu,
\]
where for every $\mu,\nu\geq 0$ 
\[
d_{\underline{l}}^{(\mu,\nu)}(x)=\sum_{j=0}^{2(l_1+\cdots +l_6)-1}a_j^{(\mu,\nu)}x^{l_1+\cdots +l_6-j}
\]
is a Laurent polynomial with coefficients $a_j^{(\mu,\nu)}$ lying in the algebra of conical sums $\mathcal{C}$, which will be defined in the next section, and $q=e^{2\pi i\tau}$.
\end{teo}

Thanks to the sum representation (\ref{sum}) it is easy to see that the functions $D_{\underline{l}}$ which are irreducible, i.e. which cannot be written as products of other two $D_{\underline{l}}$, are the ones associated with the diagrams appearing in the following figure, where any point represents a graviton:

\vspace{1 cm}
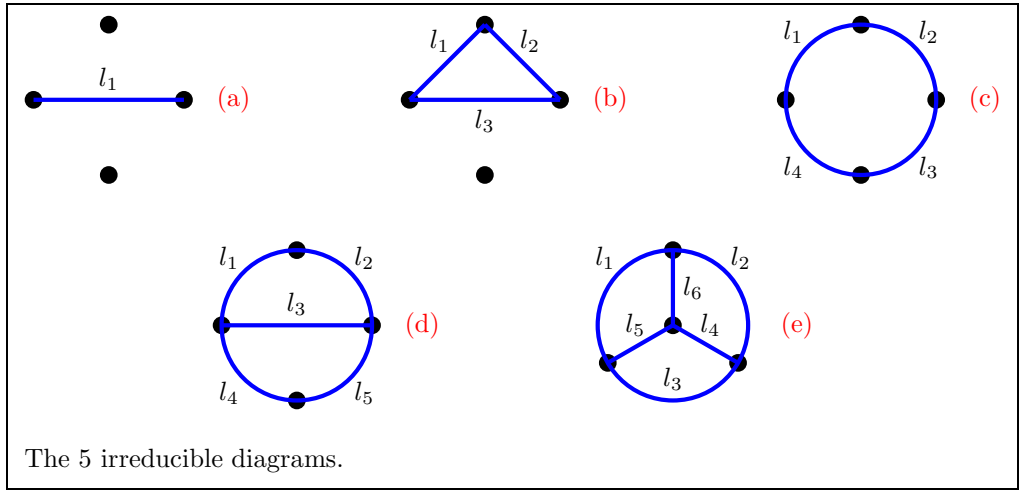
\begin{minipage}{0.26\columnwidth}
\raggedleft \fcolorbox{black}{white}
{\begin{tikzpicture}[scale=1]
\draw [fill] (1,0) circle [radius=0.1];
\draw [fill] (2,1) circle [radius=0.1];
\draw [fill] (2,-1) circle [radius=0.1];
\draw [fill] (3,0) circle [radius=0.1];
\draw [ultra thick][blue](1,0) --(3,0);
\node [above] at (2,0) {$l_1$};
\node [red, right] at (3.3,0) {(a)};
\draw [fill] (6,0) circle [radius=0.1];
\draw [fill] (7,1) circle [radius=0.1];
\draw [fill] (7,-1) circle [radius=0.1];
\draw [fill] (8,0) circle [radius=0.1];
\draw [ultra thick][blue](6,0) --(8,0);
\draw [ultra thick][blue](6,0) --(7,1);
\draw [ultra thick][blue](7,1) --(8,0);
\node [below] at (7,0) {$l_3$};
\node [above] at (6.4,0.5) {$l_1$};
\node [above] at (7.6,0.5) {$l_2$};
\node [red, right] at (8.3,0) {(b)};
\draw [fill] (11,0) circle [radius=0.1];
\draw [fill] (12,1) circle [radius=0.1];
\draw [fill] (12,-1) circle [radius=0.1];
\draw [fill] (13,0) circle [radius=0.1];
\draw [blue, ultra thick] (12,0) circle [radius=1];
\node at (11.1,0.9) {$l_1$};
\node at (12.9,0.9) {$l_2$};
\node at (12.9,-0.9) {$l_3$};
\node at (11.1,-0.9) {$l_4$};
\node [red, right] at (13.3,0) {(c)};
\draw [fill] (3.5,-3) circle [radius=0.1];
\draw [fill] (4.5,-2) circle [radius=0.1];
\draw [fill] (4.5,-4) circle [radius=0.1];
\draw [fill] (5.5,-3) circle [radius=0.1];
\draw [blue, ultra thick] (4.5,-3) circle [radius=1];
\draw [ultra thick][blue](3.5,-3) --(5.5,-3);
\node at (3.6,-2.1) {$l_1$};
\node at (5.4,-2.1) {$l_2$};
\node at (5.4,-3.9) {$l_5$};
\node at (3.6,-3.9) {$l_4$};
\node [above] at (4.5,-3) {$l_3$};
\node [red, right] at (5.8,-3) {(d)};
\draw [fill] (8.634,-3.5) circle [radius=0.1];
\draw [fill] (9.5,-2) circle [radius=0.1];
\draw [fill] (10.366,-3.5) circle [radius=0.1];
\draw [fill] (9.5,-3) circle [radius=0.1];
\draw [blue, ultra thick] (9.5,-3) circle [radius=1];
\draw [ultra thick][blue](9.5,-2) --(9.5,-3);
\draw [ultra thick][blue](8.634,-3.5) --(9.5,-3);
\draw [ultra thick][blue](10.366,-3.5) --(9.5,-3);
\node at (8.6,-2.1) {$l_1$};
\node at (10.4,-2.1) {$l_2$};
\node [above] at (9.5,-4) {$l_3$};
\node at (9,-3) {$l_5$};
\node [right] at (9.5,-2.5) {$l_6$};
\node at (10,-3) {$l_4$};
\node [red, right] at (10.8,-3) {(e)};
\node[align=left, below] at (3,-4.5)%
{The 5 irreducible diagrams.};
\end{tikzpicture}}
\end{minipage}
\vspace{1 cm}

No lines between pairs of points amount to say that there are $0$ propagators; to give an example, in the case of diagram \textcolor{red}{(a)} the associated function is $D_{l_1}:=D_{(l_1,0,0,0,0,0)}$, and we will speak of 2-point case because only 2 points are connected by a propagator.

After the introduction, in the next section, of single-valued multiple zeta values and conical sums, we will consider the 2-point case, the 3-point case (diagram \textcolor{red}{(b)}) and the 4-point case (diagrams \textcolor{red}{(c)}, \textcolor{red}{(d)} and \textcolor{red}{(e)}) separately, and we will say something about the proof of theorem~\ref{main} for all of them.\footnote{The theorem, as stated, need to be proven only in the case \textcolor{red}{(e)}, allowing the $l_i$'s to be equal to zero, and all the other cases would follow. It is however important, as we will explain, to give a more detailed proof in the simpler cases.}

The main consequence of this theorem is that, combining it with the results in section \ref{conicalsection}, we get a powerful tool to do explicit computations. In particular we focus on the 3-point case, where we go beyond what could be computed so far and we produce the first instances of multiple zeta values that cannot be reduced to products of simple zeta values. Moreover, we note that only single-valued multiple zeta values appear. To give an example, following the notation of theorem \ref{main}, we got the following Laurent polynomial in weight 7:
\begin{multline*}
d_{1,1,5}^{(0,0)}(y)=\frac{1}{4^7}\Big(\frac{62}{10945935}y^7+\frac{1}{243}\zeta_{sv}(3)y^4+\frac{119}{648}\zeta_{sv}(5)y^2 +\frac{11}{108}\zeta_{sv}(3)^2y+\frac{21}{32}\zeta_{sv}(7)\\
+\frac{23}{6}\frac{\zeta_{sv}(3)\zeta_{sv}(5)}{y}+\frac{7115\zeta_{sv}(9)-900\zeta_{sv}(3)^3}{576y^2}+\frac{1245\zeta_{sv}(3)\zeta_{sv}(7)-150\zeta_{sv}(5)^2}{64y^3}\\
+\frac{288\zeta_{sv}(3,5,3)-3600\zeta_{sv}(5){\zeta_{sv}(3)}^2-9573\zeta_{sv}(11)}{256y^4}\\
+\frac{2475\zeta_{sv}(5)\zeta_{sv}(7)+1125\zeta_{sv}(9)\zeta_{sv}(3)}{128y^5}-\frac{1575}{64}\frac{\zeta_{sv}(13)}{y^6}\Big).
\end{multline*}

\section{Conical sums}\label{conicalsection}

First let us fix the notation:
\begin{defi}
We call \emph{multiple zeta values} (MZVs) the real numbers given by the absolutely convergent series
\[
\zeta(k_1,\ldots,k_r)=\sum_{0<v_1<\cdots<v_r}\frac{1}{v_1^{k_1}\cdots v_r^{k_r}}=\sum_{\underline{x}\in\mathbb{N}^r}\frac{1}{x_1^{k_1}(x_1+x_2)^{k_2}\cdots(x_1+\cdots+ x_r)^{k_r}},
\]
where $r,k_1,\ldots,k_r\in\mathbb{N}$ and $k_r\geq 2$. We call $r$ the \emph{length} and $k:=k_1+\cdots +k_r$ the \emph{weight} of the MZV. We call $\mathcal{A}$ the algebra spanned over $\mathbb{Q}$ by the MZVs.
\end{defi}
In particular they are special values of multiple polylogarithms, which are defined for $\underline{k}=(k_1,\ldots,k_r)$ as
\[
\mbox{Li}_{\underline{k}}(z_1,\ldots,z_n)=\sum_{0<v_1<\cdots<v_r}\frac{z_1^{v_1}\cdots z_r^{v_r}}{v_1^{k_1}\cdots v_r^{k_r}}.
\]
If one considers polylogarithms in one variable (setting $z_1=\cdots =z_{n-1}=1$ and keeping only $z_n$), then one can construct their single-valued version, introduced in \cite{Brown4}. The values at~1 of these single valued functions are called \emph{single-valued multiple zeta values}. Single-valued MZVs constitute a small subset of $\mathcal{A}$, which has been introduced and studied very recently by Brown in \cite{Brown2}. We refer to that paper any time we will talk about them, in particular to section~7, where a basis is computed up to weight 13.
\begin{defi}
Let $v_1,\ldots,v_m\in\mathbb{Q}^n$, and let $\mathbb{R}^{+}$ denote the non-negative real numbers. Then we say that $C:=\mathbb{R}^{+}v_1+\cdots+\mathbb{R}^{+}v_m$ is a \emph{rational cone}, and we denote $C^{0}$ its interior.
\end{defi}
\begin{defi}
Let $C$ be a rational cone in $\mathbb{R}^n$ and let $l_1,\ldots,l_r$ be (possibly not distinct) linear forms with integer coefficients that are positive on the interior of $C$.\\
If $l_i(\underline{x})=\sum_{j=1}^n a_{i,j}x_j$, consider the matrix $A:=(a_{i,j})$. Then for $\chi$ a finite order character of~$\mathbb{Z}^n$ we define the following series:
\begin{equation}\label{eq}
\zeta(C,A,\chi):=\sum_{\underline{x}\in C^{0}\cap\mathbb{Z}^n}\frac{\chi(\underline{x})}{l_1(\underline{x})\cdots l_r(\underline{x})}
\end{equation}
When this series converges we call these numbers conical sums and we define $\mathcal{C}$ to be the vector space generated over $\mathbb{Q}$ by them. Note that $\zeta(C,A,\chi)$ does not depend on the order of the rows and columns of $A$.
\end{defi}
One can immediately see that $\mathcal{C}$ is an algebra. Setting $C$ equal to the the first quadrant in~$\mathbb{R}^2$, $l_1(\underline{x})=x_1$, $l_2(\underline{x})=l_3(\underline{x})=x_1+x_2$, and $\chi$ identically equal to 1, we get $\zeta(C,A,\chi)=\zeta(1,2)$. In the same way one gets all the MZVs, and taking non-trivial characters one gets all the other special values of polylogarithms at roots of unity.

The main reference for a systematic study of conical sums is Terasoma's paper \cite{Terasoma} (see also \cite{Paycha} for a different but less general approach to their study).

Terasoma proves that any conical sum can be reduced to a linear combination of sums of the canonical form 
\begin{equation}\label{standardcone}
\zeta(A,\chi):=\sum_{\underline{x}\in\mathbb{N}^n}\frac{\chi(\underline{x})}{l_1(\underline{x})\cdots l_r(\underline{x})},
\end{equation}
where the coefficients $a_{i,j}\in\mathbb{Z}_{\geq 0}$, and that sums of this form have the following integral representation:
\begin{equation}\label{int}
\zeta(A,\chi)=\int_{[0,1]^r}\frac{\chi(\underline{u})\,x_1^{l_1-1}\cdots x_r^{l_r-1}dx_1\cdots dx_r}{\prod_{j=1}^n(1-\chi(\underline{e}_j)\,x_1^{a_{1,j}}\cdots x_r^{a_{r,j}})},
\end{equation}
where $\underline{e}_j$ is the canonical $j$-th element of the basis of $\mathbb{Z}^n$ and $\underline{u}:=\sum_{j=1}^n \underline{e}_j=(1,\ldots,1)^T$.

This means, first of all, that all conical sums are periods. Moreover, the main result of \cite{Terasoma} is that they are linear combination of multiple polylogarithms evaluated at some roots of unity.

In particular one may be interested in understanding when a conical sum belongs to $\mathcal{A}$, and this is the case in our context, because the coefficients of the amplitudes are widely expected to be MZVs. This turns out to be a complicated problem. Once we put them in the form (\ref{standardcone}), it is clear that we should consider only sums of the kind $\zeta(A):=\zeta(A,\chi_0)$, where $\chi_0$ is the trivial character sending everything to 1.

In general, if there are coefficients in the non-negative matrix $A$ which are bigger than 1, one cannot hope to get MZVs, because this introduces a congruence condition on the sum. For example, it is an easy exercise to show that
\[
\sum_{x,y\geq 1}\frac{1}{x\,(x+2y)^2}=\frac{\pi^2\log(2)}{8}-\frac{5\zeta(3)}{16}.
\]
A natural subset that we may want to define (and that is the good set to consider in the string amplitude computations for 3 gravitons, as we will see later) is then the following:
\begin{defi}
We call $(0,1)$-matrix any matrix whose entries are only zeros and ones. Then we define $\mathcal{B}$ as the vector space generated over $\mathbb{Q}$ by the $\zeta(A)$ such that $A$ is a $(0,1)$-matrix and the sum converges.
\end{defi}
It is trivial to see that $\mathcal{B}$ is an algebra and that $\mathcal{A}\subseteq \mathcal{B}$.

It is actually bigger (assuming standard transcendence conjectures), because for instance one finds, using the method explained below, that 
\[
\sum_{x,y,z,w\geq 1}\frac{1}{(x+y)(x+y+z)(y+z+w)(x+y+w)^2}=\frac{15}{32}\zeta(5)-\frac{9}{4}\zeta(2)\zeta(3)+\frac{9}{4}\log(2)\zeta(2)^2.
\]
Using the integral representation of conical sums of the form (\ref{standardcone}) (so in particular of the numbers in $\mathcal{B}$) one can try to use the Maple program \texttt{HyperInt} recently created by E. Panzer and explained in \cite{Panzer}, which is based on ideas developed by F. Brown in \cite{Brown2006} and \cite{Brown1}. In some cases, \texttt{HyperInt} answers rewriting $\zeta(A)$ as a linear combination of multiple polylogarithms at roots of unity, and that is how we obtained the counterexample showing that $\mathcal{B}$ is strictly bigger than $\mathcal{A}$. Unfortunately, \texttt{HyperInt} up to now is not able to give an answer in general, even for numbers belonging to $\mathcal{B}$, for reasons that will not be discussed here.

Nevertheless, one can characterize\footnote{I am grateful to C. Dupont for remarking this to me, as well as for suggesting the idea of using \texttt{HyperInt} to compute conical sums.} a (not optimal) subset of numbers in $\mathcal{B}$ that will always belong to the ring $\mathcal{A}$ of MZVs (and can be computed algorithmically by \texttt{HyperInt}):
\begin{Lemma}
Let $\mathcal{S}$ be the set $\{(0,1)$-matrices such that, up to permutations of the rows, the ones are consecutive in every column$\}$. If $A\in\mathcal{S}$ and $\zeta(A)$ converges, then $\zeta(A)\in\mathcal{A}$.
\end{Lemma}
\textbf{Proof.} We can write the matrix with consecutive ones in every column, because interchanging rows does not change the sum. Recall the integral representation (\ref{int}). In our case it reduces to
\[
\zeta(A)=\int_{[0,1]^r}\frac{x_1^{l_1-1}\cdots x_r^{l_r-1}dx_1\cdots dx_r}{\prod_{j=1}^n(1-x_1^{a_{1,j}}\cdots x_r^{a_{r,j}})},
\]
where any factor in the denominator will actually be of the form $1-\prod_{k=a}^bx_k$ for $1\leq a\leq b \leq r$.
The result follows from theorem 8.2 of \cite{Brown2006}, where it is proven that integrals of this kind always belong to $\mathcal{A}$.\\
$\square$

As announced above, this condition is not optimal. Of course there are sums which we do not expect in general to be multiple zeta values, for example involving coefficients bigger than~1, that reduce to MZVs by accident (using double subdivision relations introduced in \cite{Paycha}):
\[
\sum_{x,y\geq 1} \frac{1}{(x+y)^2(2x+y)}=\frac{\zeta(3)}{4}.
\]
However, there is a set of $(0,1)$-matrices strictly bigger than $\mathcal{S}$ which gives only MZVs, an example being
\[
\sum_{x,y,z\geq 1} \frac{1}{xyz(x+y+z)},
\]
which can be easily proven by partial fraction decomposition to belong to $\mathcal{A}$ (it is equal to $6\zeta(1,1,2)$) even though its associated matrix is not in $\mathcal{S}$.

An optimal characterisation of the conical sums belonging to $\mathcal{A}$ will be the subject of a forthcoming joint work with C. Dupont.

\section{The 2-point case}

When only two gravitons are involved, with $l$ propagators between them (diagram \textcolor{red}{(a)}), the functions to compute are given by:
\[
D_{l}(\tau):=\frac{1}{4}\int G(z,\tau)^{l} \frac{dz}{\tau_2}.
\]
The behaviour of this integral as $\tau_2$ tends to infinity was already studied in \cite{Green4}. Using the same ideas we generalize that result giving the following expansion of the functions $D_l$'s, which is of course a special case of theorem \ref{main}:
\begin{teo}\label{2points}
For every $l\geq 2$
\begin{equation}\label{qqbar2}
D_l(\tau)=\sum_{\mu,\nu\geq 0}d_{l}^{(\mu,\nu)}(\pi\tau_2)q^\mu\overline{q}^\nu,
\end{equation}
where for every $\mu,\nu\geq 0$ 
\[
d_l^{(\mu,\nu)}(x)=\sum_{j=0}^{2l-1}a_j^{(\mu,\nu)}x^{l-j}
\]
is a Laurent polynomial with coefficients $a_j^{(\mu,\nu)}\in\mathcal{C}$ and $q=e^{2\pi i\tau}$.
\end{teo}
\textbf{Proof}. Using (\ref{G=B+P}) and setting $x:=z_2/\tau_2$, we want to compute
\begin{equation*}
D_{l}(\tau)=\frac{1}{2^l}\sum_{l=r+m}\frac{l!}{r!m!}\frac{(\pi\tau_2)^{r}}{2^{m}}
\iint \overline{B_2}(x)^{r}P(z,\tau)^{m} dz_1dx.
\end{equation*}
Since
\begin{equation}\label{P^m}
P(z,\tau)^{m}=\sum_{\substack{\underline{n}\in\mathbb{Z}^m\\ \underline{k}\in\mathbb{Z}_0^m}}\frac{1}{|\underline{k}|}e^{2\pi i\sum_i k_i(n_i\tau_1+z_1)}e^{-2\pi\tau_2\sum_i |k_i||n_i-x|},
\end{equation}
where $\mathbb{Z}_0:=\mathbb{Z}\setminus\{0\}$ and $|\underline{k}|:=|k_1|\cdots|k_m|$, we have
\begin{equation*}
\int_{\mathbb{R}/\mathbb{Z}} P(z,\tau)^{m} dz_1
=\sum_{\substack{\underline{n}\in\mathbb{Z}^m\\ \underline{k}\in\mathbb{Z}_0^m}}\frac{\delta_0(\underline{k})}{|\underline{k}|}e^{2\pi i(\underline{k}\cdot\underline{n})\tau_1} e^{-2\pi\tau_2(\sum|k_i||n_i-x|)},
\end{equation*}
where, for any $\underline{v}\in\mathbb{Z}^m$ and for any $a\in\mathbb{Z}$, $\delta_a(v)=1$ if the sum of the coordinates is $a$, and is zero otherwise. Moreover, $\underline{x}\cdot\underline{y}$ denotes the standard inner product of $\mathbb{R}^m$.\\
In the interval $[0,1]$,
\[
\overline{B_2}(x)^{r}=\sum_{a+b+c=r}\frac{r!}{a!\,b!\,c!}\frac{(-1)^b}{6^c}x^{2a+b}.
\]
Therefore we have
\begin{multline}\label{chaopescao}
D_l(\tau)=\\
\frac{1}{4^l}\sum_{a+b+c+m=l}\frac{l!(2\pi\tau_2)^{l-m}}{a!\,b!\,c!\,m!}\frac{(-1)^b}{6^c}\sum_{\substack{\underline{n}\in\mathbb{Z}^m\\
\underline{k}\in\mathbb{Z}_0^m}}\frac{\delta_0(\underline{k})}{|\underline{k}|}e^{2\pi i(\underline{k}\cdot\underline{n})\tau_1}\int_0^1 x^{2a+b}e^{-2\pi\tau_2(\sum|k_i||n_i-x|)} dx.
\end{multline}
To compute the last integral, let us fix $n_1,\ldots,n_m$. Since $x\in [0,1]$ we have
\[
|n_i-x|=\left\{\begin{array}{ll}
n_i-x & \mbox{if } n_i>0 \\
x-n_i & \mbox{if } n_i\leq 0,
\end{array}\right.
\]
so $\sum|k_i||n_i-x|=\sum|k_i||n_i|+x\sum\mbox{sgn}(-n_i)|k_i|$.

By repeated integration by parts one easily finds that for any $c\in\mathbb{R}_{>0}$, $\beta\in\mathbb{R}\setminus\{0\}$ and~$M\in\mathbb{N}$
\begin{equation}\label{parts}
\int_0^c x^Me^{-\beta x} dx=\frac{M!}{\beta^{M+1}}-\sum_{j=0}^M\frac{(M)_j}{\beta^{j+1}}c^{M-j}e^{-\beta c},
\end{equation}
where $(M)_j=M(M-1)\cdots (M-j+1)$ is the descending Pochhammer symbol.
In our case $\beta=2\pi\tau_2\sum\mbox{sgn}(-n_i)|k_i|$, $c=1$ and $M=2a+b$. Note that $\sum\mbox{sgn}(-n_i)|k_i|$ can be equal to zero. If it is not zero, since $|n_i|+\mbox{sgn}(-n_i)=|n_i-1|$, by (\ref{parts}) we get, as a result of the integral,
\[
\frac{(2a+b)!\;e^{-2\pi\tau_2\sum|k_i||n_i|}}{(2\pi\tau_2\sum\mbox{sgn}(-n_i)|k_i|)^{2a+b+1}}-\sum_{j=0}^{2a+b}\frac{(2a+b)!\;e^{-2\pi\tau_2\sum|k_i||n_i-1|}}{(2a+b-j)!(2\pi\tau_2\sum\mbox{sgn}(-n_i)|k_i|)^{j+1}},
\]
while if $\sum\mbox{sgn}(-n_i)|k_i|=0$ then we just get
\[
\frac{e^{-2\pi\tau_2\sum|k_i||n_i|}}{2a+b+1}.
\]
In both cases, once we fix $n_1,\ldots,n_m$ and $k_1,\ldots,k_m$, putting everything together we are left with an expression of the kind
\[
c^{(p,q)}(\pi\tau_2)e^{2\pi ip\tau_1}e^{-2\pi q\tau_2},
\]
where $p=\sum k_in_i \in\mathbb{Z}$, $q$ is a non-negative integer equal to either $\sum|k_i||n_i-1|$ or $\sum|k_i||n_i|$, and $c^{(p,q)}(\pi\tau_2)$ is a Laurent polynomial with rational (explicitly determined) coefficients whose maximum power is $l$ and minimum power is $1-l$.

Note that $e^{2\pi ip\tau_1}e^{-2\pi q\tau_2}=q^\mu\overline{q}^\nu$, with $\mu=(q+p)/2$ and $\nu=(q-p)/2$. Therefore we would like to show that $q\geq |p|$ and that $p\equiv_2 q$ in order to have that $\mu$ and $\nu$ are non-negative integers. If $q=\sum|k_i||n_i|$ the claim is trivial, so we have to take care only of the case $q=\sum|k_i||n_i-1|$.

Note that $k_in_i \equiv_2 |k_i||n_i| \equiv_2 |k_i|(1+|n_i-1|)$ and that $\sum |k_i| \equiv_2 \sum k_i=0$, so
\[
p=\sum k_in_i \equiv_2 \sum|k_i| + \sum|k_i||n_i-1| \equiv_2 q.
\]
Moreover $|p|=|\sum k_in_i|+|\sum k_i|\leq |\sum k_i(n_i-1)|\leq \sum|k_i||n_i-1|=q$. 

To conclude our proof we have to analyse more carefully the rational coefficients of $\tilde{c}^{(\mu,\nu)}(\pi\tau_2)$, which are obtained by the $c^{(p,q)}(\pi\tau_2)$'s.

Any fixed $q=\sum|k_i||n_i-\varepsilon|$ ($\varepsilon=0,1$) can be obtained with just finitely many $m$-tuples $(n_1,\ldots,n_m)$, because $|n_i-\varepsilon|\leq q$ for any $i$: otherwise, since for every $i$ $|k_i|\geq 1$, we would have $\sum |k_i||n_i-\varepsilon|\geq \sum |n_i-\varepsilon|\geq q$.
This means that for any $(\mu,\nu)$, which is uniquely determined by a couple $(p,q)$, one has to consider a finite rational linear combination of sums of the kind
\[
T(m,\underline{n},\alpha):=\sum_{\underline{k}}\frac{\delta_0(\underline{k})}{|\underline{k}|(\sum\mbox{sgn}(-n_i)|k_i|)^\alpha},
\]
where $\underline{k}\in\mathbb{Z}_0^m$ are such that $\sum\mbox{sgn}(-n_i)|k_i|\neq 0$ and $\alpha$ is a positive integer. Note that the function $T$ only depends on $n_i/|n_i|$.

Since we can split the sum defining $T$ as a sum over cones such that the forms in the denominator are either bigger or smaller than zero, it follows that our coefficients are linear combinations of conical sums.\\
$\square$

Specializing carefully this computation to the $(p,q)=(0,0)$ case, one gets the result already found in \cite{Green4}:
\begin{cor}
\begin{eqnarray*}
d_{l}(x):=d^{(0,0)}_{l}(x)&=&\Big(\frac{x}{12}\Big)^{l} {}_2F_1(1,-l,3/2;3/2)\\
&+&\frac{2}{4^l}\sum_{\substack{a+b+c+m=l\\m\geq 2}}\frac{l!(2a+b)!}{a!b!c!m!}\frac{(-1)^b}{6^c}S(m,2a+b+1)(2x)^{c-a-1},
\end{eqnarray*}
where
\[
{}_2F_1(1,-l,3/2;3/2)=\sum_{j=0}^l\frac{l!(-1)^j}{(l-j)!(3/2)_n}\Big(\frac{3}{2}\Big)^n
\]
is the classical Gauss hypergeometric function and $S(m,\alpha):=T(m,(1,\ldots,1),\alpha)$.
\end{cor}
\textbf{Proof.} Note that since $q=\sum|k_i||n_i-\varepsilon|$ with $\varepsilon=0$ or $1$ the only possible $m$-tuples $(n_1,\ldots,n_m)$ which can give $q=0$ are $(0,\ldots,0)$ and $(1,\ldots 1)$, and with them also $p=0$, because $\sum k_i=0$. Using this and looking carefully\footnote{As explained in \cite{Green4}, one exploits the fact that $\overline{B}_2(1-x)=\overline{B}_2(x)$ in order to get the nice looking formula in the corollary instead of the more complicated one then we would naively get just by performing the same steps of the theorem's proof.} at the proof of the previous theorem one is then lead to the formula above, except perhaps for the hypergeometric coefficient of the leading term, which is a bit fancier than what one obtains with the integration process described in the proof, and can be deduced by making use of the identity
\[
\frac{1}{x(x+1)\cdots (x+n)}=\sum_{j=0}^n \frac{(-1)^j}{j!(n-j)!}\frac{1}{x+j},
\]
easily obtained by partial fraction decomposition.\\
$\square$

The function $S(m,\alpha)$ was proven by D. Zagier in \cite{Zagier2} to be equal to an explicit linear combination of MZVs, allowing to algorithmically compute in terms of MZVs the non-exponentially small part of $D_l$.

Noting a few mistakes in the data given in \cite{Green4}, the coefficients of these Laurent polynomials always happen to be polynomials in simple odd zeta-values, as it is also the case in the computation of the coefficients of the genus zero 4-point amplitude (see \cite{Green7}).

This is not surprising: one can think of the limit when $\tau_2$ is big as a degeneracy of the torus with 2 marked points to a sphere with 4 marked points, and Zagier recently managed, finding a way to write the coefficients of $d_{l}$ in terms of the coefficients of the 4-point amplitude in genus zero, to prove the following [Zagier, unpublished]:
\begin{teo}\label{zagier}
For all integers $l$ the Laurent polynomial $d_l(\pi\tau_2)$ has coefficients belonging to the polynomial ring generated over $\mathbb{Q}$ by the simple odd zeta values $\zeta(2n+1)$.
\end{teo}

Unfortunately, knowing $d_{l}$ is not enough to perform the integration over the moduli space of complex tori, so one would like to understand better the behaviour of the functions $D_l$. This can be achieved by looking at the more general theorem \ref{2points}, because it allows us to predict other coefficients of the expansion (\ref{qqbar2}).

To make an example, let us recall that the non-holomorphic Eisenstein series are defined, for $s\in\mathbb{C}$ with $\Re(s)>1$ and $\tau\in\mathbb{H}$, by
\begin{equation}\label{eisenstein}
E_s(\tau)={\sum_{(m,n)\in\mathbb{Z}^2\setminus\{(0,0)\}}} \frac{\tau_2^s}{|m\tau +n|^{2s}}.
\end{equation}
We actually want to consider the modified function $E_s^\prime(\tau):=(4\pi)^{-s}E_s(\tau)$. It is a real analytic function, modular invariant with respect to the action of $\mbox{SL}_2(\mathbb{Z})$. If we suppose that $s=n\in\mathbb{N}$, $n\geq 2$, its explicit expansion at the cusp is given, setting $y=\pi\tau_2$, by
\begin{multline}\label{eisensteincusp}
E_n^{\prime}(\tau)= \frac{1}{4^n}\Big[(-1)^{n-1}\frac{B_{2n}}{(2n)!}(4y)^n+\frac{4(2n-3)!}{(n-2)!(n-1)!}\zeta(2n-1)(4y)^{1-n}\\
+\frac{2}{(n-1)!}\sum_{N\geq 1}N^{n-1}\sigma_{1-2n}(N)(q^N+\overline{q}^N)\sum_{m=0}^{n-1}\frac{(n+m-1)!}{m!(n-m-1)!}(4Ny)^{-m}\Big],
\end{multline}
where $B_n$ is the $n$-th Bernoulli number, and $\sigma_k(N)=\sum_{d|N}d^k$ is a finite power sum running over the positive divisors of $N$.

Using the series representations (\ref{sum}) for $D_l(\tau)$ and (\ref{eisenstein}) for the Eisenstein series one can immediately see that $D_2(\tau)=E_2^{\prime}(\tau)$.

Let us briefly see how can we get the same result by comparing the expansion given by theorem \ref{2points} and the expansion (\ref{eisensteincusp}), which becomes in this case
\[
E_2^{\prime}(\tau)=\frac{1}{16}\Big[\frac{y^2}{45}+\frac{\zeta(3)}{y}+2\sum_{N\geq 1}(2N+y^{-1})\sigma_{-3}(N)(q^N+\overline{q}^N)\Big].
\]
Using the intermediate step (\ref{chaopescao}) in the proof of the theorem, for $m=0$ we get $(1/16)(y^2/45)$, which is the leading term of the non-exponential part, for $m=1$ we get $0$, and for $m=2$ (so $a=b=c=0$) we get
\[
\frac{1}{16}\sum_{\substack{k\in\mathbb{Z}\setminus\{0\}\\ n\in\mathbb{Z}}}\frac{e^{2\pi ip\tau_1}e^{-2\pi\tau_2(|k|(|n_1|+|n_2|))}}{|k|^2}\int_0^1 e^{-2\pi\tau_2x|k|(\text{sgn}(-n_1)+\text{sgn}(-n_2))} dx,
\]
where we denote $p=k(n_1-n_2)$.
When $p=0$, i.e. when $q$ and $\overline{q}$  have the same power in the expansion given by the theorem, then $n_1=n_2$, which implies that the argument of the exponential in the integral is never zero. Therefore, splitting the sum into the $n\geq 1$ part and the $n\leq 0$ part, one gets 2 telescoping sums, both giving as a result
\[
\sum_{k\in\mathbb{Z}\setminus\{0\}}\frac{1}{64|k|^3y}.
\]
We conclude that the variable $\tau_1$ does not appear only in the non-exponentially small part of $D_2(\tau)$, which is the Laurent polynomial $d_2(y)=(y^2/45+\zeta(3)/y)/16$. This fits with the expansion of $E_2^{\prime}(\tau)$.

Moreover, if $p>0$ one gets again telescoping sums in $n$, but now we sum only over finitely many $k$, which are the divisors of $p$. We leave as an exercise to the reader to verify that one gets exactly the same expansion as we get for the non-holomorphic Eisenstein series.

In general it is too messy to repeat the same game as above and explicitly get the full expansion for other $D_l$'s, except maybe for $D_3(\tau)$, which is already known to be equal to $E_3^{\prime}(\tau)+\zeta(3)$ (Zagier, unpublished. Recently an alternative simpler proof was given in \cite{Green5}). It is possible, however, to algorithmically get the coefficient of $q^\mu\overline{q}^\nu$ for any fixed $(\mu,\nu)$, which in principle allows to check conjectures on the full expansion or to numerically approximate the functions very precisely (the sum (\ref{qqbar2}) converges much faster than the sum~(\ref{sum})).

\section{The 3-point case}

When three particles are involved the only new irreducible diagram that we have to consider is diagram \textcolor{red}{(b)}, with all $l_i$'s strictly positive, whose associated $D_{\underline{l}}$ reads:
\[
D_{\underline{l}}(\tau):=\frac{1}{4^3}\iint G(z,\tau)^{l_1}G(w,\tau)^{l_2}G(z-w,\tau)^{l_3} \frac{dzdw}{\tau_2^2},
\]
where the integration is performed over 2 copies of the torus associated with the lattice~$\Lambda_\tau$.

We now give the proof of theorem \ref{main} also for this case, because it helps to understand how to explicitly get the coefficients. Since the ideas used here are exactly the same as in the previous section, and the notation gets much heavier, we will give less details. It is however important to understand how the generalization to this case exploits the same ideas used for 2 particles, because in the next section we will give only a sketch of the proof in the general case, assuming that one has already understood how to take care of the missing details.
\begin{teo}\label{trepunti}
For every $\underline{l}=(l_1,l_2,l_3)$ we have
\[
D_{\underline{l}}(\tau)=\sum_{\mu,\nu\geq 0}d_{\underline{l}}^{(\mu,\nu)}(\pi\tau_2)q^\mu\overline{q}^\nu,
\]
where for every $\mu,\nu\geq 0$ 
\[
d_{\underline{l}}^{(\mu,\nu)}(x)=\sum_{j=0}^{2(l_1+l_2+l_3)-1}a_j^{(\mu,\nu)}x^{l_1+l_2+l_3-j}
\]
is a Laurent polynomial with coefficients $a_j^{(\mu,\nu)}\in\mathcal{C}$ and $q=e^{2\pi i\tau}$.
\end{teo}
\textbf{Proof.} Let us introduce the following notations: $\underline{l}!:=l_1!\,l_2!\,l_3!$, and $c\,^{\underline{l}}:=c\,^{l_1+l_2+l_3}$. Moreover, for $\underline{k}=(k_1,...,k_m)$ we write $|\underline{k}|:=|k_1|\cdots |k_m|$, and $\Vert\underline{k}\Vert:=|k_1|+\cdots +|k_m|$.\\
With the substitutions $x=z_2/\tau_2$, $y=w_2/\tau_2$ we get
\begin{multline*}
D_{\underline{l}}(\tau)=\frac{1}{2^{\underline{l}}}\sum_{\underline{r}+\underline{m}=\underline{l}}\frac{\underline{l}!}{\underline{r}!\,\underline{m}!}\frac{(\pi\tau_2)^{\underline{r}}}{2^{\underline{m}}}\times \\
\times \iint \overline{B_2}(x)^{r_1}\overline{B_2}(y)^{r_2}\overline{B_2}(x-y)^{r_3}P(z,\tau)^{m_1}P(w,\tau)^{m_2}P(z-w,\tau)^{m_3} dz_1dw_1dxdy.
\end{multline*}

Using (\ref{P^m}) we have
\begin{multline*}
\iint_{(\mathbb{R}/\mathbb{Z})^{2}} P(z,\tau)^{m_1}P(w,\tau)^{m_2}P(z-w,\tau)^{m_3} dz_1dw_1 = \\
=\sum_{\substack{(\underline{k},\underline{h},\underline{t})\\(\underline{n},\underline{p},\underline{q})}}\frac{\delta_0((\underline{k},\underline{h}))\delta_0((\underline{k},\underline{t}))}{|\underline{k}|\,|\underline{h}|\,|\underline{t}|}e^{2\pi i(\underline{k}\cdot\underline{n}+\underline{h}\cdot\underline{p}+\underline{t}\cdot\underline{q})\tau_1} e^{-2\pi\tau_2(\sum|k_i||n_i-x|+\sum|h_i||p_i-y|+\sum|t_i||q_i-(x-y)|)},
\end{multline*}
where the sum runs over $(\underline{k},\underline{h},\underline{t})\in\mathbb{Z}_0^{m_1}\times\mathbb{Z}_0^{m_2}\times\mathbb{Z}_0^{m_3}$ and $(\underline{n},\underline{p},\underline{q})\in\mathbb{Z}^{m_1}\times\mathbb{Z}^{m_2}\times\mathbb{Z}^{m_3}$.

Then we need to calculate, for any fixed $\underline{r}$, $\underline{m}$, $(\underline{k},\underline{h},\underline{t})$ and $(\underline{n},\underline{p},\underline{q})$,
\begin{equation*}
\iint_{[0,1]^2}B_2(x)^{r_1}B_2(y)^{r_2}B_2(x-y)^{r_3}e^{-2\pi\tau_2(\sum|k_i||n_i-x|+\sum|h_i||p_i-y|+\sum|t_i||q_i-(x-y)|)}dxdy.
\end{equation*}
Since $\sum|k_i||n_i-x|=\sum|k_i||n_i|+x\sum\mbox{sgn}(-n_i)|k_i|$, this is equal to
\begin{equation*}
e^{-2\pi\tau_2(\sum |k_i||n_i|+\sum |h_i||p_i|+\sum |t_i||q_i|)}\iint_{[0,1]^2}B_2(x)^{r_1}B_2(y)^{r_2}B_2(x-y)^{r_3}e^{-\gamma x}e^{-\delta y}dxdy,
\end{equation*}
where $\gamma:=2\pi\tau_2(\sum\mbox{sgn}(-n_i)|k_i|+\sum\mbox{sgn}(-q_i)|t_i|)$ and $\delta:=2\pi\tau_2(\sum\mbox{sgn}(-p_i)|h_i|+\sum\mbox{sgn}(q_i)|t_i|)$.

This is equal to
\begin{multline}\label{dueint}
\sum_{\underline{a}+\underline{b}+\underline{c}=\underline{r}}\frac{\underline{r}!}{\underline{a}!\,\underline{b}!\,\underline{c}!}\frac{(-1)^{\underline{b}}}{6^{\underline{c}}}e^{-2\pi\tau_2(\sum |k_i||n_i|+\sum |h_i||p_i|+\sum |t_i||q_i|)}\times \\
\times \Big(\int_0^1 x^{2a_1+b_1}e^{-\gamma x} dx\int_0^x y^{2a_2+b_2}(x-y)^{2a_3+b_3}e^{-\delta y} dy \\
+\int_0^1 y^{2a_2+b_2}e^{-\delta y} dy\int_0^y x^{2a_1+b_1}(y-x)^{2a_3+b_3}e^{-\gamma x} dx\Big).
\end{multline}
Since the integral that are left to compute are completely specular, let us describe just the result of the first one. After using the binomial theorem on $(x-y)^{2a_3+b_3}$, we get a $\mathbb{Q}$-linear combination of integrals of the kind
\[
\int_0^1 x^{M}e^{-\gamma x} dx\int_0^1 y^{N}e^{-\delta y} dy.
\]
Now, as in the 2-point case, we use integration by parts and finally get a linear combination of $1$, $e^{-(\gamma+\delta)}$ and $e^{-\gamma}$, with coefficients that are products of polynomials in $\delta^{-1}$  and $(\gamma~+~\delta)^{-1}$ with rational coefficients (with some obvious modifications in case $\gamma$ and/or $\delta$ are zero). We do not give the exact formula here, for reasons of space, except for the special case of the non-exponentially small term, which we describe in the next corollary.
However, it is easy to see, going through the computation, that for any fixed $(\underline{k},\underline{h},\underline{t})$ and $(\underline{n},\underline{p},\underline{q})$ we get a term of the kind $c^{(p,q)}(\pi\tau_2)e^{2\pi ip\tau_1}e^{-2\pi q\tau_2}$, where $p=\sum k_in_i+\sum h_ip_i+\sum t_iq_i \in\mathbb{Z}$, $q$ is a non-negative integer equal to $\sum|k_i||n_i-1|+\sum|h_i||p_i-1|+\sum|t_i||q_i-1|+\vartheta$, with $\vartheta=0$ or $\sum\mbox{sgn}(-n_i)|k_i|+\sum\mbox{sgn}(-q_i)|t_i|$ or $\sum\mbox{sgn}(-n_i)|k_i|+\sum\mbox{sgn}(-h_i)|q_i|$ or $\sum\mbox{sgn}(-p_i)|h_i|+\sum\mbox{sgn}(q_i)|t_i|$, and $c^{(p,q)}(\pi\tau_2)$ is a Laurent polynomial with rational (explicitly determined) coefficients, whose maximum power is $l_1+l_2+l_3$ and minimum power is $1-(l_1+l_2+l_3)$.

For all the possible $\vartheta$ we can apply the method described in the 2-point case to prove that $q\geq |p|$ and that $p\equiv_2 q$. Moreover, again one can prove that for any $(\mu,\nu)$ only finitely many $(\underline{n},\underline{p},\underline{q})$ are allowed, and the coefficients of the Laurent polynomials belong to $\mathcal{C}$ (and are very explicitly determined).\\
$\square$

In particular one can deduce the following (already found in \cite{Green4}):
\begin{cor}
\[
d_{\underline{l}}(x):=4^{\underline{l}}\;d^{(0,0)}_{\underline{l}}(x)=d_{\underline{l}}^{A}(x)+d_{\underline{l}}^{B}(x)+d_{\underline{l}}^{C}(x)
\]
and the three contributions are defined as follows:
\begin{equation*}
d_{\underline{l}}^{A}(x)=2(2x)^{\underline{l}}\sum_{\underline{a}+\underline{b}+\underline{c}=\underline{l}}\frac{\underline{l}!}{\underline{a}!\underline{b}!\underline{c}!}\frac{(-1)^{\underline{b}}}{6^{\underline{c}}}\frac{(2a_2+b_2)!(2a_3+b_3)!}{(2(a_2+a_3)+b_2+b_3+1)!}\frac{1}{\lambda+1}
\end{equation*}
is the contribution for $m_1=m_2=m_3=0$, where $\lambda:=2(a_1+a_2+a_3)+b_1+b_2+b_3+1$;
\[
d_{\underline{l}}^B(x)=d^B_{l_1,l_2,l_3}(x)+d^B_{l_2,l_1,l_3}(x)+d^B_{l_3,l_2,l_1}(x),
\]
with
\begin{multline*}
d^B_{l_1,l_2,l_3}(x)=\sum_{\substack{\underline{a}+\underline{b}+\underline{c}+\underline{m}=\underline{l}\\u+v=2a_3+b_3\\e+f=2a_1+b_1+u}} 2\frac{\underline{l}!}{\underline{a}!\underline{b}!\underline{c}!\underline{m}!}\frac{(-1)^{\underline{b}}}{6^{\underline{c}}}(2x)^{\underline{c}-\underline{a}-2}\times \\
\times \frac{(-1)^{v}(2a_3+b_3)!(2a_1+b_1+u)!(2a_2+b_2+v+f)!}{u!v!f!}R(m_1,m_2,m_3;2a_2+b_2+v+f+1,e+1),
\end{multline*}
is the contribution when at least 2 of the $m_i$'s are $>0$, with
\[
R(m_1,m_2,m_3;\alpha ,\beta ):=\sum_{(\underline{k},\underline{h},\underline{t})} \frac{\delta_0((\underline{k},\underline{h}))\delta_0((\underline{k},\underline{t}))}{|\underline{k}||\underline{h}||\underline{t}|(\Vert\underline{k}\Vert + \Vert\underline{h}\Vert)^\alpha(\Vert\underline{k}\Vert + \Vert\underline{t}\Vert)^\beta};
\]
\[
d_{\underline{l}}^C(x)=d^C_{l_1,l_2,l_3}(x)+d^C_{l_2,l_1,l_3}(x)+d^C_{l_3,l_1,l_2}(x),
\]
with
\begin{multline*}
d^C_{l_1,l_2,l_3}(x)=\sum_{\substack{\underline{a}+\underline{b}+\underline{c}+\underline{m}=\underline{l}\\u+v=2a_3+b_3}} 2\frac{\underline{l}!}{\underline{a}!\,\underline{b}!\,\underline{c}!\,\underline{m}!}\frac{(-1)^{\underline{b}}}{6^{\underline{c}}}(2x)^{\underline{c}-\underline{a}-2}\times \\
\times \frac{(2a_3+b_3)!}{u!\,v!}\frac{(-1)^{v}}{2a_2+b_2+v+1}(\lambda)!\times\\
\times \Big[S(m_1,\lambda+1)+ \sum_{j=0}^{\lambda}\frac{(-1)^j S(m_1,j+1)}{(\lambda-j)!}(2\pi\tau_2)^{(\lambda-j)}\Big],
\end{multline*}
is the contribution when 2 of the $m_i$'s are zero ($m_2$ and $m_3$ in $d^C_{l_1,l_2,l_3}$), where again $\lambda:=2(a_1+a_2+a_3)+b_1+b_2+b_3+1$ and 
$S(m,\alpha)=R(m,0,0,\alpha,0)$ is the sum already introduced in the previous section. In every case where at least one of the $m_i$'s in $d^B$ and $d^C$ is zero, we assume that the other $m_i$'s are $\geq 2$, otherwise it is easy to see that the contribution given is zero.
\end{cor}
Note that we are computing the non-exponentially small term without dividing it by $4^{\underline{l}}$, in order to have neater results afterwards. This does not change the proof, but may generate some confusion concerning the resulting expression.

\textbf{Sketch of the proof.} It is convenient to consider as separated cases: the one with all the $m_i$'s equal zero; the one with 2 of the $m_i$'s equal zero and one bigger than $1$ (it cannot be equal to one!); the one with one of the $m_i$'s equal zero and $2$ bigger than $1$; the one with all of them bigger than zero. The last case is the most complicated (and it gives the same result as the case when only one is zero); we briefly describe how to treat it, and the same argument can be applied to the other cases.

Following the proof of the theorem above one arrives at the point (\ref{dueint}), and then should take into account only the $(\underline{n},\underline{p},\underline{q})$ leading to non-exponentially small terms. These are $(\underline{0},\underline{0},\underline{0})$, $(\underline{1},\underline{0},\underline{1})$ and $(\underline{1},\underline{1},\underline{0})$ for the first integral, $(\underline{1},\underline{1},\underline{0})$, $(\underline{0},\underline{1},\underline{-1})$ and $(\underline{0},\underline{0},\underline{0})$ for the second integral. By substituting $x=1-x$ and $y=1-y$ in the second integral one gets just 2 copies of the~3 possible cases for the first integral. We call the first one $d^B_{l_1,l_2,l_3}(x)$, and then one can easily notice that the other 2 are given by $d^B_{l_2,l_1,l_3}(x)$ and $d^B_{l_3,l_2,l_1}(x)$.\\
$\square$

Let us remark that, since by definition $D_{\underline{l}}$ does not depend on the order of the $l_i$'s, also $d_{\underline{l}}$ does not, even though from this formula it is not clear at first sight. So, for instance, speaking of $d_{1,2,3}$ is the same as speaking of $d_{3,1,2}$.

Evaluating by hands the functions $R(m_1,m_2,m_3;\alpha ,\beta )$ in terms of MZVs, one is able to find for the lower weights:
\begin{eqnarray*}
d_{1,1,1}(y)&=&\frac{2}{945}y^3+\frac{3}{4}\frac{\zeta(5)}{y^2},\\
d_{1,1,2}(y)&=&\frac{2}{14175}y^4+\frac{\zeta(3)}{45}y +\frac{5}{12}\frac{\zeta(5)}{y}- \frac{1}{4}\frac{\zeta(3)^2}{y^2}+\frac{9}{16}\frac{\zeta(7)}{y^3},\\
d_{1,1,3}(y)&=&\frac{2}{22275}y^5+\frac{\zeta(3)}{45}y^2+\frac{11}{60}\zeta(5)+\frac{105}{32}\frac{\zeta(7)}{y^2}-\frac{3}{2}\frac{\zeta(3)\zeta(5)}{y^3}+\frac{81}{64}\frac{\zeta(9)}{y^4},\\
d_{1,2,2}(y)&=&\frac{8}{467775}y^5+\frac{4\zeta(3)}{945}y^2+\frac{13}{45}\zeta(5)+\frac{7}{8}\frac{\zeta(7)}{y^2}-\frac{\zeta(3)\zeta(5)}{y^3}+\frac{9}{8}\frac{\zeta(9)}{y^4}.
\end{eqnarray*}

These are all the possible cases up to weight 5. The same result was found in \cite{Green5}, noting some mistakes in \cite{Green4}. 

Note that until this weight the coefficients of the Laurent polynomials are MZVs, but in the literature it is not proven yet that this will happen in any weight. Moreover let us remark that they are MZVs of a very particular kind: they are always polynomials in simple odd zeta values, as well as it happens (and is proven to be so in any weight by theorem \ref{zagier}) in the 2-point case. What we are now able to say is that, using theorem \ref{trepunti} and Terasoma's result, they have to be special values at roots of unity of multiple polylogarithms.

For higher weights the sums $R(m_1,m_2,m_3;\alpha ,\beta )$ look impossible to be evaluated by hands, hence no other $d_{\underline{l}}$ was known so far.

Since we have seen that the coefficients are conical sums, one can hope to use \texttt{HyperInt} if the cones and the matrices involved are simple enough. This turns out to be the case with 3 particles, because of the following theorem:
\begin{teo}\label{th1}
For all $l_1,l_2,l_3\in\mathbb{N}$ the coefficients of $d_{l_1,l_2,l_3}(y)$ belong to the algebra $\mathcal{B}$.
\end{teo}
The proof of this theorem is constructive, and gives an actual formula to compute the coefficients, but the formula itself is very long and complicated. It can be found in the \emph{Appendix} (see equations \ref{zero},...,\ref{otto}), together with the proof.

Thus one gets an algorithm which will certainly compute the coefficients of $d_{\underline{l}}$ in terms of MZVs for any $\underline{l}$ such that only matrices lying in $\mathcal{S}$ are involved. For example all the $d_{\underline{l}}$ of weight 6 satisfy this condition, after some partial fraction decomposition on the conical sums, but not all the weight 7: we will come back to this later. Moreover, as remarked before, the algorithm will produce an answer, either in terms of MZVs or alternating sums, for many more $(0,1)$-matrices than just the ones in $\mathcal{S}$.

Here come the new data obtained so far with this method (we set again $y:=\pi\tau_2$):
\begin{eqnarray*}
d_{1,1,4}(y)&=&\frac{284}{18243225}y^6+\frac{2}{135}\zeta(3)y^3+\frac{5\zeta(5)}{18}y +\frac{1}{10}\zeta(3)^2+\frac{51}{20}\frac{\zeta(7)}{y}+\frac{11}{2}\frac{\zeta(5)\zeta(3)}{y^2}\\&+&\frac{79\zeta(9)-36\zeta(3)^3}{24y^3}-\frac{9}{4}\frac{\zeta(3)\zeta(7)}{y^4}+\frac{45}{16}\frac{\zeta(11)}{y^5},\\
d_{2,2,2}(y)&=&\frac{193}{11609325}y^6+\frac{1}{315}\zeta(3)y^3+\frac{59}{315}\zeta(5)y+\frac{23}{20}\frac{\zeta(7)}{y}+\frac{5}{2}\frac{\zeta(3)\zeta(5)}{y^2}-\frac{65}{48}\frac{\zeta(9)}{y^3}\\&+&\frac{21\zeta(5)^2-18\zeta(3)\zeta(7)}{16y^4}+\frac{99}{64}\frac{\zeta(11)}{y^5},\\
d_{1,2,3}(y)&=&\frac{298}{42567525}y^6+\frac{1}{315}\zeta(3)y^3+\frac{173}{1260}\zeta(5)y +\frac{3}{20}\zeta(3)^2+\frac{53}{20}\frac{\zeta(7)}{y}-\frac{5}{2}\frac{\zeta(3)\zeta(5)}{y^2}\\&+&\frac{223\zeta(9)+96\zeta(3)^3}{32y^3}-\frac{99\zeta(5)^2+162\zeta(3)\zeta(7)}{32y^4}+\frac{729}{128}\frac{\zeta(11)}{y^5},
\end{eqnarray*}
\begin{multline*}
d_{1,1,5}(y)=\frac{62}{10945935}y^7+\frac{2}{243}\zeta(3)y^4+\frac{119}{324}\zeta(5)y^2 +\frac{11}{27}\zeta(3)^2y+\frac{21}{16}\zeta(7)\\
+\frac{46}{3}\frac{\zeta(3)\zeta(5)}{y}+\frac{7115\zeta(9)-3600\zeta(3)^3}{288y^2}+\frac{1245\zeta(3)\zeta(7)-150\zeta(5)^2}{16y^3}\\
+\frac{288\zeta(3,5,3)-288\zeta(3)\zeta(3,5)-5040\zeta(5)\zeta(3)^2-9573\zeta(11)}{128y^4}\\
+\frac{2475\zeta(5)\zeta(7)+1125\zeta(9)\zeta(3)}{32y^5}-\frac{1575}{32}\frac{\zeta(13)}{y^6},
\end{multline*}
\begin{multline*}
d_{1,3,3}(y)=\frac{34}{8513505}y^7+\frac{2}{945}\zeta(3)y^4+\frac{17}{252}\zeta(5) y^2+\frac{23}{105}\zeta(3)^2y+\frac{1391}{560}\zeta(7)\\
-\frac{3\zeta(3)\zeta(5)}{y} + \frac{953\zeta(9)+144\zeta(3)^3}{32y^2}-\frac{1701\zeta(3)\zeta(7)+120\zeta(5)^2}{32y^3}\\
+\frac{324\zeta(3,5,3)-324\zeta(3)\zeta(3,5)+22299\zeta(11)+8460\zeta(5)\zeta(3)^2}{320y^4}\\
-\frac{891\zeta(5)\zeta(7)+702\zeta(9)\zeta(3)}{16y^5}+\frac{7209}{128}\frac{\zeta(13)}{y^6},  
\end{multline*}
\begin{multline*}
d_{1,1,6}(y)=\frac{262}{186080895}y^8+\frac{1}{243}\zeta(3)y^5+\frac{113}{324}\zeta(5)y^3+\frac{25}{36}\zeta(3)^2y^2+\frac{749}{144}\zeta(7)y
+\frac{331}{18}\zeta(3)\zeta(5)\\
+\frac{56\zeta(9)-207\zeta(3)^3}{18y}+\frac{705\zeta(3)\zeta(7)+375\zeta(5)^2}{2y^2}\\
+\frac{2304\zeta(3,5,3)-2304\zeta(3)\zeta(3,5)-38541\zeta(11)-32400\zeta(5)\zeta(3)^2}{64y^3}+\frac{a}{y^4}\\
+\frac{b}{y^5}+\frac{179550\zeta(11)\zeta(3)+274050\zeta(9)\zeta(5)+155925\zeta(7)^2}{128y^6}-\frac{1233225}{512}\frac{\zeta(15)}{y^7},
\end{multline*}
where $a$ cannot be determined because of current limits of \texttt{HyperInt}, and
\begin{multline*}
b=\frac{837}{14}\zeta(5)\zeta(5,3)-\frac{3375}{4}\zeta(3){\zeta(5)}^2-\frac{6075}{8}\zeta(7){\zeta(3)}^2-\frac{675}{56}\zeta(3,7,3)\\+\frac{675}{56}\zeta(3)\zeta(7,3)+\frac{54}{7}\zeta(5,3,5)+\frac{135}{4}\zeta(5)\zeta(8)-\frac{134257}{896}\zeta(13).
\end{multline*}
Starting from weight 7, we can see something new and very interesting happening to the coefficients:
not only polynomials in odd simple zeta values are involved. For example the coefficient of $y^{-4}$ in $d_{1,1,5}(y)$ contains $\zeta(3,5,3)$ and $\zeta(3)\zeta(3,5)$, which are not reducible to polynomials in odd zetas.
The fundamental remark is that they are still very special, because that coefficient can be written as the following linear combination of single-valued multiple zeta values:
\[
\frac{9}{8}\zeta_{sv}(3,5,3)-\frac{225}{16}\zeta_{sv}(5){\zeta_{sv}(3)}^2-\frac{9573}{256}\zeta_{sv}(11).
\]
This actually happens to all of the coefficients in the polynomials above (products of odd zeta values are already single-valued MZVs), the most astonishing case being the coefficient of $y^{-5}$ in $d_{1,1,6}(y)$, that we called $b$. Indeed, one can check that, in terms of the basis for single valued MZVs in weight 13 in \cite{Brown2},
\[
b=\frac{27}{7}\zeta_{sv}(5,3,5)-\frac{675}{112}\zeta_{sv}(3,7,3)-\frac{4995}{4}\zeta_{sv}(3){\zeta_{sv}(5)}^2-\frac{7425}{8}\zeta_{sv}(7){\zeta_{sv}(3)}^2-\frac{134257}{1792}\zeta_{sv}(13).
\]
This means that a multiple zeta value a priori belonging to a vector space of dimension 16 actually belongs to the subspace of dimension 5 of single-valued MZVs, which looks much more than a coincidence.
Let us now draw a parallel between our setting and the closed strings tree-level: when the genus is zero, in the most trivial case (4 particles) only odd zetas appear, but going to the next case (5 particles) in general one finds also MZVs of bigger length, and the coefficients belong to the algebra of single valued MZVs. Therefore we conclude that it is not too optimistic to conjecture that the coefficients are given by single valued MZVs only, after such a little evidence. Arguments supporting this conjecture, based on the structure of the one-loop string amplitude, are given in the forthcoming paper~\cite{Green8}.

We conclude this section by observing that, unfortunately, the matrices appearing do not always belong to $\mathcal{S}$, even after performing standard manipulations like partial fractioning, and sometimes they produce polylogarithms at higher roots of unity. This happens, for instance, in the computation of $R(3,3,1;1,1)$. However, in this case we only get alternating sums, which are good enough to be computed by \texttt{HyperInt}, and in the end all the non-MZV part of $R(3,3,1;1,1)$ cancels out.

It is actually very tempting to conjecture that the numbers $R(m_1,m_2,m_3;\alpha,\beta)$ themselves always lie in $\mathcal{A}$ (but they are not single-valued), since our data confirm it so far, however we do not have any other argument this time to support this evidence.

A very partial result in the direction of proving the conjectures above is the following:
\begin{teo}
For any $n\in\mathbb{N}$ the coefficients of $d_{1,1,n}(\tau)$ are linear combination of conical sums whose matrices belong to $\mathcal{S}$, so in particular they are (algorithmically) $\mathbb{Q}$-linear combinations of multiple zeta values.
\end{teo}
\textbf{Proof.} The proof uses the explicit formula given in the \emph{Appendix} for the numbers
$R$ in terms of elements of $\mathcal{B}$. The only $R$'s involved are of the kind $R(1,1,j;\alpha,\beta)$, $R(1,j,1;\alpha,\beta)$ and $R(j,1,1;\alpha,\beta)$, with $j\leq n$ and some $\alpha$, $\beta$.
Note that $R(1,j,1;\alpha,\beta)=R(1,1,j;\beta,\alpha)$ and $R(j,1,1;\alpha,\beta)=R(1,1,j;0,\alpha+\beta)$, so it is enough to study $R(1,1,j;\alpha,\beta)$. Only the sums (\ref{uno}) and (\ref{quattro}) are contributing to this $R$, but the sum (\ref{uno}) is easily seen to be contained in $\mathcal{A}$, so we have to study (\ref{quattro}) only, which in our case is particularly simple (assume $j\geq 2$, otherwise (\ref{uno}) suffices):
\[
\sum_{\substack{Q,F\geq 0\\Q+F=j-2}}\sum_{\substack{l_3+a>q_1>\cdots >q_Q\\l_3>f_1>\cdots >f_F}}\frac{1}{{(l_3+a)}^{\beta+1}a^{\alpha+2}l_3q_1\cdots q_Qf_1\cdots f_F}
\]
Following the stuffle procedures described in the \emph{Appendix} one is left with a linear combination of sums in $\mathcal{B}$ with associated matrices of the kind
\[ 
\begin{pmatrix}
  1 &  &  &  &  &  &  \\
  \vdots & \ddots &  &  &  &  &  \\
  1  & \cdots  & 1 &  &  &  &  \\
    &  &  & 1 &  &  &  \\
    & A &  & \vdots & \ddots &  &  \\
    &  &  & 1 & \cdots & 1 &  \\
  1 & \cdots & 1 & 0 & \cdots & 0 & 1 \\
  1 & \cdots & 1 & 1 & \cdots & 1 & 1 
 \end{pmatrix}
\]
where $A$ is a matrix with rows given either by consecutive ones or by consecutive zeros (this comes from the stuffle).
Since interchanging the rows does not change the conical sum, we can rewrite the matrix as
\[ 
\begin{pmatrix}
  1 &  &  &  &  &  &  \\
  \vdots & \ddots &  &  &  &  &  \\
  1  & \cdots  & 1 &  &  &  &  \\
  1 & \cdots & 1 &  &  &  & 0 \\
  \vdots &  & \vdots &  & B &  & \vdots \\
  1 & \cdots & 1 &  &  &  & 0 \\
  1 & \cdots & 1 & 1 & \cdots & 1 & 1 \\
  1 & \cdots & 1 & 0 & \cdots & 0 & 1 \\
  0 & \cdots & 0 &  &  &  & 0 \\
  \vdots &  & \vdots &  & C &  & \vdots \\
  0 & \cdots & 0 &  &  &  & 0 
 \end{pmatrix}
\]
where $B$ and $C$ are matrices with, from left to right, a string of ones followed by a string of zeros in every row, such that the length of the string of ones increases in $B$ with the increase of the row's index and decreases in $C$. At this point we almost have a matrix belonging to $\mathcal{S}$, the only problem being the row in the middle of the form $r=1,\ldots,1,0,\ldots,0,1$.

Note now that a partial fraction operation on the sum of the kind
\[
\frac{1}{l_i(\underline{x})\,l_j(\underline{x})}=\frac{1}{l_i(\underline{x})\,(l_i+l_j)(\underline{x})}+\frac{1}{l_j(\underline{x})\,(l_i+l_j)(\underline{x})}
\]
is reflected on the matrix just by substituting the $i$-th or the $j$-th row by the sum of the 2. Hence if we do this sum operation on $r$ together with the row immediately below we get the sum of 2 matrices, one belonging to $\mathcal{S}$ (when $r$ is deleted) and one such that the sub-matrix below $r$ is strictly smaller (after interchanging $r$ with the new row obtained as a sum). Iterating this process one finally gets that $r$ is the last row in the matrix, and in this case the matrix belongs to $\mathcal{S}$ and we are done.\\
$\square$

\section{The 4-point case}

In this section we will briefly explain why the techniques seen in details in the previous sections allow us to prove theorem \ref{main} in its more general statement, with 4 gravitons interacting on a torus.

\textbf{Proof of theorem \ref{main} (sketch).} The most difficult part of proving this result, after having proved the 2-point case and 3-point case, is probably to find an acceptable notation. Recall that we are dealing with the integral
\[
D_{\underline{l}}(\tau)=\frac{1}{4^3}\iint_{(\mathbb{C}/\Lambda_\tau)^3} G(z-w,\tau)^{l_1}G(v,\tau)^{l_2}G(w-v,\tau)^{l_3}G(z,\tau)^{l_4}G(z-v,\tau)^{l_5}G(w,\tau)^{l_6} \frac{dzdwdv}{\tau_2^3}
\]
\begin{multline*}
=\frac{1}{2^{\underline{l}}}\sum_{\underline{r}+\underline{m}=\underline{l}}\frac{\underline{l}!}{\underline{r}!\,\underline{m}!}\frac{(\pi\tau_2)^{\underline{r}}}{2^{\underline{m}}}\iint \overline{B_2}(x-y)^{r_1}\overline{B_2}(u)^{r_2}\overline{B_2}(y-u)^{r_3}\overline{B_2}(x)^{r_4}\overline{B_2}(x-u)^{r_5}\overline{B_2}(y)^{r_6}\times\\
\times P(z-w,\tau)^{m_1}P(v,\tau)^{m_2}P(w-v,\tau)^{m_3}P(z,\tau)^{m_4}P(z-v,\tau)^{m_5}P(w,\tau)^{m_6} dzdwdv,
\end{multline*}
where $x=z_2/\tau_2$, $y=w_2/\tau_2$ and $u=v_2/\tau_2$. The integration over $\tau_1$ gives
\begin{multline*}
\sum \frac{\delta_0((\underline{k}^{(1)},\underline{k}^{(4)},\underline{k}^{(5)}))\delta_0((-\underline{k}^{(1)},\underline{k}^{(3)},\underline{k}^{(6)}))\delta_0((\underline{k}^{(2)},-\underline{k}^{(3)},-\underline{k}^{(5)}))}{|\underline{k}^{(1)}|\cdots |\underline{k}^{(6)}|}\prod_{j=1}^6 e^{2\pi i\tau_1(\sum k_i^{(j)}n_i^{(j)})}\\
\times \int_{[0,1]^3} \overline{B_2}(x-y)^{r_1}\cdots \overline{B_2}(y)^{r_6} e^{-2\pi\tau_2(\sum |k_i^{(1)}||n_i^{(1)}-(x-y)|)}\cdots e^{-2\pi\tau_2(\sum |k_i^{(6)}||n_i^{(6)}-y|)},
\end{multline*}
where the sum runs over $\underline{k}^{(j)}\in\mathbb{Z}_0^{m_j}$ and $\underline{n}^{(j)}\in\mathbb{Z}^{m_j}$ for $j=1,\ldots,6$. Let us call $I$ the integral appearing in the last step. We have
\begin{multline}\label{peacenlove}
I=\sum_{\underline{a}+\underline{b}+\underline{c}=\underline{r}}\frac{\underline{r}!}{\underline{a}!\underline{b}!\underline{c}!}\frac{(-1)^{\underline{b}}}{6^{\underline{c}}}\prod_{j=1}^6 e^{-2\pi\tau_2(\sum |k_i^{(j)}||n_i^{(j)}|)}\times\\
\times \Big(\int_{P_1}+\cdots +\int_{P_6}\Big)(|x-y|^{M_1}u^{M_2}|y-u|^{M_3}x^{M_4}|x-u|^{M_5}y^{M_6}e^{-\gamma_1 x}e^{-\gamma_2 y}e^{-\gamma_3 u} dxdydu),
\end{multline}
where the $P_i$ is the path $0\leq\sigma_i(u)\leq\sigma_i(y)\leq\sigma_i(x)\leq 1$, for $\sigma_i$ a permutation of the 3 variables $x$, $y$, $u$, $M_i:=2a_i+b_i$, and the $\gamma_i$'s will depend on the path chosen.

Let us consider the path $P_1$ with $\sigma_1=\mbox{Id}$ only ($0\leq v\leq y\leq x\leq 1$). There we have
\[
\gamma_1:=2\pi\tau_2 (\sum\mbox{sgn}(-n_i^{(1)})|k_i^{(1)}|+\sum\mbox{sgn}(-n_i^{(4)})|k_i^{(4)}|+\sum\mbox{sgn}(-n_i^{(5)})|k_i^{(5)}|),
\]
\[
\gamma_2:=2\pi\tau_2 (\sum\mbox{sgn}(n_i^{(1)})|k_i^{(1)}|+\sum\mbox{sgn}(-n_i^{(3)})|k_i^{(3)}|+\sum\mbox{sgn}(-n_i^{(6)})|k_i^{(6)}|),
\]
\[
\gamma_3:=2\pi\tau_2 (\sum\mbox{sgn}(-n_i^{(2)})|k_i^{(2)}|+\sum\mbox{sgn}(n_i^{(3)})|k_i^{(3)}|+\sum\mbox{sgn}(n_i^{(5)})|k_i^{(5)}|).
\]
The integral on this path reduces to a linear combination of integrals of the kind
\[
\int_0^1 x^{N_1}e^{-\gamma_1 x} dx \int_0^x y^{N_2}e^{-\gamma_2 y} dy \int_0^y u^{N_3}e^{-\gamma_3 u} du,
\]
where the $N_i$'s are non negative integers. One can solve the integral by repeatedly using integration by parts, and all the possible exponentials involved in the result are $e^{-\gamma_1}$, $e^{-(\gamma_1+\gamma_2)}$ and $e^{-(\gamma_1+\gamma_2+\gamma_3)}$. Multiplying them by the exponential in front of the integral in formula (\ref{peacenlove}) tell us what are all the possible integers $q$ in the terms of the kind $e^{-2\pi q\tau_2}$.

It is not difficult to see that the argument used in the 2-point case works for all of these $q$'s, and that nothing new happens to the Laurent polynomials involved and to their coefficients, which are therefore expressible as conical sums.\\
$\square$

We do not write down here an explicit formula for the Laurent polynomial part of the functions $D_{\underline{l}}(\tau)$, because it is really complicated and does not really allow one to work with it. Indeed, the same method explained in the previous section to explicitly write down the conical sums as integrals produces, in the 4-point case, matrices with coefficients strictly bigger than~1, whose computation in terms of special values of polylogarithms goes beyond the nowadays limits of \texttt{HyperInt}. 

\section*{Acknowledgements}

I would like to thank Don Zagier, Clément Dupont, Erik Panzer, Pierre Vanhove, Michael B. Green, Oliver Schlotterer, Nils Matthes, Johannes Brödel, Stephan Stieberger and Alessandro Valentino for useful suggestions, comments, discussions and advices. I would like to thank Erik Panzer also for providing me with a version of \texttt{HyperInt} which could perform much more efficiently the computations needed.

My research was supported by the Max Planck Institute for Mathematics.

\section*{Appendix}

\textbf{Proof of theorem \ref{th1}.} In the beginning we will partially exploit the same ideas (and notations) of \cite{Zagier2}, so we will be slightly sketchy, referring the reader to that reference for more details. 

Let us recall the definition of the function $R$:
\[
R(m_1,m_2,m_3;\alpha ,\beta ):=\sum_{(\underline{k},\underline{h},\underline{t})} \frac{\delta_0((\underline{k},\underline{h}))\delta_0((\underline{k},\underline{t}))}{|\underline{k}||\underline{h}||\underline{t}|{(\Vert\underline{k}\Vert + \Vert\underline{h}\Vert)}^\alpha{(\Vert\underline{k}\Vert + \Vert\underline{t}\Vert)}^\beta}.
\]
Note that if we have $\sum_ik_i=a$ for some $a\in\mathbb{Z}$, then we impose, using the condition in the numerator, that also $\sum_ih_i=a$ and $\sum_it_i=a$. This means that we can rewrite the series as $R(m_1,m_2,m_3;\alpha ,\beta )=R_0(m_1,m_2,m_3;\alpha ,\beta )+2R_{>0}(m_1,m_2,m_3;\alpha ,\beta )$ where
\begin{equation*}
\label{S_0}
R_0(m_1,m_2,m_3;\alpha ,\beta ):=\sum_{(\underline{k},\underline{h},\underline{t})} \frac{\delta_0(\underline{k})\delta_0(\underline{h})\delta_0(\underline{t})}{|\underline{k}||\underline{h}||\underline{t}|{(\Vert\underline{k}\Vert + \Vert\underline{h}\Vert)}^\alpha{(\Vert\underline{k}\Vert + \Vert\underline{t}\Vert)}^\beta}
\end{equation*}
and 
\begin{equation*}
R_{>0}(m_1,m_2,m_3;\alpha ,\beta ):=\sum_{a\geq 1}\sum_{(\underline{k},\underline{h},\underline{t})} \frac{\delta_a(\underline{k})\delta_a(\underline{h})\delta_a(\underline{t})}{|\underline{k}||\underline{h}||\underline{t}|{(\Vert\underline{k}\Vert + \Vert\underline{h}\Vert)}^\alpha{(\Vert\underline{k}\Vert + \Vert\underline{t}\Vert)}^\beta}.
\end{equation*}
We define, for $l\geq 1$ and for $r\geq 0$,
\[
S_{r}(l):=\sum_{\substack{k_1,\ldots,k_r\geq 1\\k_1+\cdots+k_r=l}}\frac{1}{|\underline{k}|},
\]
setting $S_{r}(l)=0$ if $r=0$ or if $r>l$.

Let us consider now $R_0(m_1,m_2,m_3;\alpha ,\beta )$: if $r_1$ is the number of positive $k_i$'s, $r_2$ and $r_3$ the same for the $h_i$'s and the $t_i$'s, and $l_1$ (resp. $l_2$ and $l_3$) is the sum of the positive $k_i$'s (resp. $h_i$'s and $t_i$'s), then
\begin{eqnarray*}
R_0(m_1,m_2,m_3;\alpha ,\beta )&=& \sum_{\substack{r_1=0,\ldots ,m_1\\r_2=0,\ldots ,m_2\\r_3=0,\ldots ,m_3}}\sum_{l_1,l_2,l_3\geq 1}\frac{\prod_{i=1}^3\binom{m_i}{r_i}S_{r_i}(l_i)S_{m_i-r_i}(l_i)}{{(2l_1+2l_2)}^\alpha {(2l_1+2l_3)}^\beta }\\
&=&\frac{1}{2^{\alpha+\beta}}\sum_{l_1,l_2,l_3\geq 1}\frac{\prod_{i=1}^3\mbox{coeff}_{x^{l_i}y^{l_i}}\big[\big(\mbox{Li}_1(x)+\mbox{Li}_1(y)\big)^{m_i}\big]}{{(l_1+l_2)}^\alpha{(l_1+l_3)}^\beta},
\end{eqnarray*}
where $\mbox{Li}_1(x)=\sum_{k\geq 1}x^k/k$.

Hence we get the generating function
\begin{multline*}
\sum_{m_1,m_2,m_3\geq 0}\frac{R_0(m_1,m_2,m_3;\alpha ,\beta )}{m_1!m_2!m_3!}X^{m_1}Y^{m_2}Z^{m_3}=\frac{1}{2^{\alpha+\beta}}\sum_{l_1,l_2,l_3\geq 1}\frac{\binom{X+l_1-1}{l_1}^2\binom{Y+l_2-1}{l_2}^2\binom{Z+l_3-1}{l_3}^2}{{(l_1+l_2)}^\alpha{(l_1+l_3)}^\beta}\\
=\frac{1}{2^{\alpha+\beta}}\sum_{l_1,l_2,l_3\geq 1}\frac{{X}^2{Y}^2{Z}^2}{{(l_1+l_2)}^\alpha{(l_1+l_3)}^\beta l_1^2l_2^2l_3^2}\prod_{n=1}^{l_1-1}\Big(1+\frac{X}{n}\Big)^2\prod_{p=1}^{l_2-1}\Big(1+\frac{Y}{p}\Big)^2\prod_{q=1}^{l_3-1}\Big(1+\frac{Z}{q}\Big)^2
\end{multline*}
\begin{multline*}
=\frac{1}{2^{\alpha+\beta}}\sum_{l_1,l_2,l_3\geq 1}\frac{{X}^2{Y}^2{Z}^2}{{(l_1+l_2)}^\alpha{(l_1+l_3)}^\beta l_1^2l_2^2l_3^2}\times \\
\times\prod_{n=1}^{l_1-1}\Big(1+4\sum_{\gamma\in\{1,2\}}\frac{(X/2)^\gamma}{n^\gamma}\Big)\prod_{p=1}^{l_2-1}\Big(1+4\sum_{\delta\in\{1,2\}}\frac{(Y/2)^\delta}{p^\delta}\Big)\prod_{q=1}^{l_3-1}\Big(1+4\sum_{\varepsilon\in\{1,2\}}\frac{(Z/2)^\varepsilon}{q^\varepsilon}\Big)\\
=\frac{1}{2^{\alpha+\beta}}\sum_{N,P,Q\geq 1}{\sum}^\prime\frac{4^{N+P+Q}{X}^{2+\underline{\gamma}}{Y}^{2+\underline{\delta}}{Z}^{2+\underline{\varepsilon}}}{2^{\underline{\gamma}+\underline{\delta}+\underline{\varepsilon}}{(l_1+l_2)}^\alpha{(l_1+l_3)}^\beta l_1^2l_2^2l_3^2n_1^{\gamma_1}\cdots n_N^{\gamma_N}p_1^{\delta_1}\cdots p_P^{\delta_P}q_1^{\varepsilon_1}\cdots q_Q^{\varepsilon_Q}},
\end{multline*}
where the sum ${\sum}^\prime$ runs over $l_1>n_1>\cdots>n_N>0$, $l_2>p_1>\cdots>p_P>0$, $l_3>q_1>\cdots>q_Q>0$ and over all the $\gamma_i$, $\delta_i$ and $\varepsilon_i$ belonging to $\{1,2\}$.

Comparing the coefficients we get
\begin{multline}\label{zero}
R_0(m_1,m_2,m_3;\alpha ,\beta )=\\
\frac{m_1!m_2!m_3!}{2^{\alpha+\beta+m_1+m_2+m_3-6}}{\sum}^{\prime\prime}\frac{2^{2(N+P+Q)}}{{(l_1+l_2)}^\alpha{(l_1+l_3)}^\beta l_1^2l_2^2l_3^2n_1^{\gamma_1}\cdots n_N^{\gamma_N}p_1^{\delta_1}\cdots p_P^{\delta_P}q_1^{\varepsilon_1}\cdots q_Q^{\varepsilon_Q}},
\end{multline}
where ${\sum}^{\prime\prime}$ runs over all the $N,P,Q$, and over all the $\gamma_i$, $\delta_i$ and $\varepsilon_i$ belonging to $\{1,2\}$ such that $\gamma_1+\cdots+\gamma_N=m_1-2$, $\delta_1+\cdots+\delta_P=m_2-2$, $\varepsilon_1+\cdots+\varepsilon_Q=m_3-2$, as well as over $l_1>n_1>\cdots>n_N>0$, $l_2>p_1>\cdots>p_P>0$, $l_3>q_1>\cdots>q_Q>0$.
Note that this sum is not zero only if all the $m_i$'s are strictly bigger than $1$. Note also that, by definition of $\mathcal{B}$, $R_0(m_1,m_2,m_3;\alpha ,\beta )\in\mathcal{B}$.

Now let us consider the more complicated sum $R_{>0}(m_1,m_2,m_3;\alpha ,\beta )$:
\begin{eqnarray*}
R_{>0}(m_1,m_2,m_3;\alpha ,\beta )&=& \sum_{a\geq 1}\sum_{\substack{r_1=0,\ldots ,m_1\\r_2=0,\ldots ,m_2\\r_3=0,\ldots ,m_3}}\sum_{l_1,l_2,l_3\geq 1}\frac{\prod_{i=1}^3\binom{m_i}{r_i}S_{r_i}(l_i+a)S_{m_i-r_i}(l_i)}{{(2l_1+2l_2+2a)}^\alpha {(2l_1+2l_3+2a)}^\beta }\\
&=&\frac{1}{2^{\alpha+\beta}}\sum_{\substack{l_1,l_2,l_3\geq 1\\a\geq 1}}\frac{\prod_{i=1}^3\mbox{coeff}_{x^{l_i+a}y^{l_i}}\big[\big(\mbox{Li}_1(x)+\mbox{Li}_1(y)\big)^{m_i}\big]}{{(l_1+l_2+a)}^\alpha{(l_1+l_3+a)}^\beta},
\end{eqnarray*}
Therefore we find the generating function
\begin{equation*}
2^{\alpha+\beta}\sum_{m_1,m_2,m_3\geq 0}\frac{R_{>0}(m_1,m_2,m_3;\alpha ,\beta )}{m_1!m_2!m_3!}X^{m_1}Y^{m_2}Z^{m_3}
\end{equation*}
\begin{eqnarray*}
&=&\sum_{\substack{l_1,l_2,l_3\geq 0\\a\geq 1}}\frac{\binom{X+l_1+a-1}{l_1+a}\binom{X+l_1-1}{l_1}\binom{Y+l_2+a-1}{l_2+a}\binom{Y+l_2-1}{l_2}\binom{Z+l_3+a-1}{l_3+a}\binom{Z+l_3-1}{l_3}}{{(l_1+l_2+a)}^\alpha{(l_1+l_3+a)}^\beta}\\
&=&\sum_{a\geq 1}\frac{\binom{X+a-1}{a}\binom{Y+a-1}{a}\binom{Z+a-1}{a}}{a^{\alpha+\beta}}\\
&+&\sum_{l_1,a\geq 1}\frac{\binom{X+l_1+a-1}{l_1+a}\binom{X+l_1-1}{l_1}\binom{Y+a-1}{a}\binom{Z+a-1}{a}}{{(l_1+a)}^{\alpha+\beta}}\\
&+&\sum_{l_2,a\geq 1}\frac{\binom{X+a-1}{a}\binom{Y+l_2+a-1}{l_2+a}\binom{Y+l_2-1}{l_2}\binom{Z+a-1}{a}}{{(l_2+a)}^{\alpha}{a}^\beta}\\
&+&\sum_{l_3,a\geq 1}\frac{\binom{X+a-1}{a}\binom{Y+a-1}{a}\binom{Z+l_3+a-1}{l_3+a}\binom{Z+l_3-1}{l_3}}{{(l_3+a)}^{\beta}{a}^\alpha}
\end{eqnarray*}
\begin{eqnarray*}
&+&\sum_{l_1,l_2,a\geq 1}\frac{\binom{X+l_1+a-1}{l_1+a}\binom{X+l_1-1}{l_1}\binom{Y+l_2+a-1}{l_2+a}\binom{Y+l_2-1}{l_2}\binom{Z+a-1}{a}}{{(l_1+l_2+a)}^\alpha{(l_1+a)}^\beta}\\
&+&\sum_{l_1,l_3,a\geq 1}\frac{\binom{X+l_1+a-1}{l_1+a}\binom{X+l_1-1}{l_1}\binom{Y+a-1}{a}\binom{Z+l_3+a-1}{l_3+a}\binom{Z+l_3-1}{l_3}}{{(l_1+a)}^\alpha{(l_1+l_3+a)}^\beta}\\
&+&\sum_{l_2,l_3,a\geq 1}\frac{\binom{X+a-1}{a}\binom{Y+l_2+a-1}{l_2+a}\binom{Y+l_2-1}{l_2}\binom{Z+l_3+a-1}{l_3+a}\binom{Z+l_3-1}{l_3}}{{(l_2+a)}^\alpha{(l_3+a)}^\beta}\\
&+&\sum_{l_1,l_2,l_3,a\geq 1}\frac{\binom{X+l_1+a-1}{l_1+a}\binom{X+l_1-1}{l_1}\binom{Y+l_2+a-1}{l_2+a}\binom{Y+l_2-1}{l_2}\binom{Z+l_3+a-1}{l_3+a}\binom{Z+l_3-1}{l_3}}{{(l_1+l_2+a)}^\alpha{(l_1+l_3+a)}^\beta}
\end{eqnarray*}
The idea is to apply to all these sums the same method shown for $R_0(m_1,m_2,m_3;\alpha ,\beta )$. Just to fix the notation, we write down explicitly what happens with the last and most complicated sum:
\begin{equation*}
\sum_{l_1,l_2,l_3,a\geq 1}\frac{\binom{X+l_1+a-1}{l_1+a}\binom{X+l_1-1}{l_1}\binom{Y+l_2+a-1}{l_2+a}\binom{Y+l_2-1}{l_2}\binom{Z+l_3+a-1}{l_3+a}\binom{Z+l_3-1}{l_3}}{{(l_1+l_2+a)}^\alpha{(l_1+l_3+a)}^\beta}
\end{equation*}
\begin{eqnarray*}
&=&\sum_{l_1,l_2,l_3,a\geq 1}\frac{{X}^2{Y}^2{Z}^2}{{(l_1+l_2+a)}^\alpha{(l_1+l_3+a)}^\beta(l_1+a)(l_2+a)(l_3+a)l_1l_2l_3}\times\\
&\times & \prod_{n=1}^{l_1+a-1}\Big(1+\frac{X}{n}\Big)\prod_{d=1}^{l_1-1}\Big(1+\frac{X}{d}\Big)\prod_{p=1}^{l_2+a-1}\Big(1+\frac{Y}{p}\Big)\prod_{e=1}^{l_2-1}\Big(1+\frac{Y}{e}\Big)\prod_{q=1}^{l_3+a-1}\Big(1+\frac{Z}{q}\Big)\prod_{f=1}^{l_3-1}\Big(1+\frac{Z}{f}\Big)\\
&=&\sum_{\substack{N,D\geq 0\\P,E\geq 0\\Q,F\geq 0}}{\sum}^{\sim}\frac{{X}^{2+N+D}{Y}^{2+P+E}{Z}^{2+Q+F}}{{(l_1+l_2+a)}^\alpha{(l_1+l_3+a)}^\beta(l_1+a)(l_2+a)(l_3+a)l_1l_2l_3n_1\cdots f_F}
\end{eqnarray*}
where ${\sum}^{\sim}$ runs over $a\geq 1$, $l_1+a>n_1>\cdots>n_N>0$, $l_1>d_1>\cdots>d_D>0$, $l_2+a> p_1>\cdots>p_P>0$, $l_2>e_1>\cdots>e_E>0$, $l_3+a>q_1>\cdots>q_Q>0$, $l_3>f_1>\cdots>f_F>0$.

Doing this for all the sums involved and comparing the coefficients, one finally obtains that $(m_1!m_2!m_3!/2^{\alpha+\beta})R_{>0}(m_1,m_2,m_3;\alpha ,\beta )$ is
\begin{eqnarray}
\label{uno}&=&{\sum}\frac{1}{a^{\alpha+\beta+3}n_1\cdots q_{m_3-1}}\\
\label{due}&+&\sum_{\substack{N,D\geq 0\\N+D=m_1-2}}{\sum}\frac{1}{{(l_1+a)}^{\alpha+\beta+1}a^2l_1n_1\cdots q_{m_3-1}}\\
\label{tre}&+&\sum_{\substack{P,E\geq 0\\P+E=m_2-2}}{\sum}\frac{1}{{(l_2+a)}^{\alpha+1}a^{\beta+2}l_2n_1\cdots q_{m_3-1}}\\
\label{quattro}&+&\sum_{\substack{Q,F\geq 0\\Q+F=m_3-2}}{\sum}\frac{1}{{(l_3+a)}^{\beta+1}a^{\alpha+2}l_3n_1\cdots f_F}\\
\label{cinque}&+&\sum_{\substack{N,D,P,E\geq 0\\N+D=m_1-2\\P+E=m_2-2}}{\sum}\frac{1}{{(l_1+l_2+a)}^{\alpha}{(l_1+a)}^{\beta+1}(l_2+a)a\;l_1l_2n_1\cdots q_{m_3-1}}
\end{eqnarray}
\begin{eqnarray}
\label{sei}&+&\sum_{\substack{N,D,Q,F\geq 0\\N+D=m_1-2\\Q+F=m_3-2}}{\sum}\frac{1}{{(l_1+a)}^{\alpha+1}{(l_1+l_3+a)}^{\beta}(l_3+a)a\;l_1l_3n_1\cdots f_F}\\
\label{sette}&+&\sum_{\substack{P,E,Q,F\geq 0\\P+E=m_2-2\\Q+F=m_3-2}}{\sum}\frac{1}{{(l_2+a)}^{\alpha+1}{(l_3+a)}^{\beta+1}a\;l_2l_3n_1\cdots f_F}\\
\label{otto}&+&\sum_{\substack{N,D,P,E\geq 0\\N+D=m_1-2\\P+E=m_2-2\\Q+F=m_3-2}}{\sum}\frac{1}{{(l_1+l_2+a)}^\alpha{(l_1+l_3+a)}^\beta(l_1+a)(l_2+a)(l_3+a)l_1l_2l_3n_1\cdots f_F}
\end{eqnarray}
The sum in (\ref{uno}) runs over $n_1>\cdots >n_{m_1-1}>0$, $p_1>\cdots >p_{m_2-1}>0$, $q_1>\cdots >q_{m_3-1}>0$, $a>\max\{n_1,p_1,q_1\}$.\\
The sum in (\ref{due}) runs over $l_1+a>n_1>\cdots >n_N>0$, $l_1>d_1>\cdots >d_D>0$, $p_1>\cdots >p_{m_2-1}>0$, $q_1>\cdots >q_{m_3-1}>0$, $a>\max\{p_1,q_1\}$, and is $0$ if $m_1=1$.\\
The sum in (\ref{tre}) runs over $n_1>\cdots >n_{m_1-1}>0$, $l_2+a>p_1>\cdots >p_P>0$, $l_2>e_1>\cdots >e_E>0$, $q_1>\cdots >q_{m_3-1}>0$, $a>\max\{n_1,q_1\}$, and is $0$ if $m_2=1$.\\
The sum in (\ref{quattro}) runs over $n_1>\cdots >n_{m_1-1}>0$, $p_1>\cdots >p_{m_2-1}>0$, $l_3+a>q_1>\cdots >q_Q>0$, $l_3>f_1>\cdots >f_F>0$, $a>\max\{n_1,p_1\}$, and is $0$ if $m_3=1$.\\
The sum in (\ref{cinque}) runs over $l_1+a>n_1>\cdots >n_N>0$, $l_1>d_1>\cdots >d_D>0$, $l_2+a>p_1>\cdots >p_P>0$, $l_2>e_1>\cdots >e_E>0$, $a>q_1>\cdots >q_{m_3-1}>0$, and is $0$ if $m_1=1$ or $m_2=1$.\\
The sum in (\ref{sei}) runs over $l_1+a>n_1>\cdots >n_N>0$, $l_1>d_1>\cdots >d_D>0$, $a>p_1>\cdots >p_{m_2-1}>0$, $l_3+a>q_1>\cdots >q_Q>0$, $l_3>f_1>\cdots >f_F>0$, and is $0$ if $m_1=1$ or $m_3=1$.\\
The sum in (\ref{sette}) runs over $a>n_1>\cdots >n_{m_1-1}>0$, $l_2+a>p_1>\cdots >p_P>0$, $l_2>e_1>\cdots >e_E>0$, $l_3+a>q_1>\cdots >q_Q>0$, $l_3>f_1>\cdots >f_F>0$, and is $0$ if $m_2=1$ or $m_3=1$.\\
The sum in (\ref{otto}) runs over $l_1+a>n_1>\cdots >n_N>0$, $l_1>d_1>\cdots >d_D>0$, $l_2+a>p_1>\cdots >p_P>0$,  $l_2>e_1>\cdots >e_E>0$, $l_3+a>q_1>\cdots >q_Q>0$, $l_3>f_1>\cdots >f_F>0$, and is $0$ if one of the $m_i$'s is $<2$.

From this formula it is not clear yet whether these numbers are in $\mathcal{B}$, so one needs to quasi-shuffle, or stuffle, some groups of variables.

In (\ref{uno}) one has to stuffle the 3 groups of ordered variables $n_i$, $p_i$, $q_i$; then setting $a>\max\{n_1,p_1,q_1\}$ we directly get MZV.\\
In (\ref{due}) one has to stuffle the 2 groups of ordered variables $p_i$, $q_i$ and the 2 groups of ordered variables $n_i$ and $l_1>d_1>\cdots >d_D$, in order to get sums of the kind, for $1\leq i\leq N$ and $N,M\geq 1$,
\[
\sum_{\substack{y_M>\cdots >y_1>0 \\ x_i+y_M>x_N>\cdots >x_1>0}}\frac{1}{x_1^{\eta_1}\cdots y_M^{\eta_{N+M}} (x_i+y_M)^\varepsilon}.
\]
Furthermore, if we stuffle the groups of ordered variables $y_M>\cdots >y_1>0$ and $x_N-x_i>\cdots >x_{i+1}-x_i>0$, then we get numbers in $\mathcal{B}$.\\
The same reasoning works with some obvious modification for all the other sums, and this proves our assertion.\\
$\square$

\bibliographystyle{plain}

\markboth{\textsc{Bibliography}}{\textsc{Bibliography}}
\vspace{4,1 cm}
\textsc{Max Planck Institut F\"ur Mathematik}\\
\textsc{Vivatsgasse 7, 53115, Bonn, Germany}\\

\textsc{fzerbini@mpim-bonn.mpg.de}
\end{document}